\documentclass[useAMS,usenatbib]{mn2e}
\usepackage{hyperref}
\usepackage{times}
\usepackage{natbib}
\usepackage{aas_macros}
\usepackage{amsmath}
\usepackage{amssymb}
\usepackage[figuresright]{rotating}
\usepackage[labelfont=bf]{caption}
\usepackage{supertabular}
\usepackage{xcolor}

\newcommand{\ergs}{$\mathrm{erg\;s^{-1}}$}

\newcommand{\sar}{\lambda_\mathrm{sBHAR}}
\newcommand{\Sar}{$\sar$}
\newcommand{\msun}{\mathcal{M}_\odot}

\newcommand{\mstel}{\mathcal{M}_*}
\newcommand{\Mstel}{$\mstel$}
\newcommand{\lx}{L_\mathrm{X}}
\newcommand{\LX}{$\lx$}
\newcommand{\giv}{\;|\;}
\newcommand{\Psar}{$p(\sar)$}

\newcommand{\sfratio}{$\mathrm{SFR/SFR_{MS}}(z)$}
\newcommand{\sfr}{\mathrm{SFR}}

\newcommand{\avsar}{\langle \sar \rangle}
\newcommand{\Avsar}{$\avsar$}
\newcommand{\fduty}{f(\sar>0.01)}
\newcommand{\Fduty}{$\fduty$}
\newcommand{\fbright}{f(\sar>0.1)}
\newcommand{\Fbright}{$\fbright$}
\newcommand{\Pbs}{$p($BHAR/SFR$)$}

\defcitealias{Aird17}{Paper~I} 
\newcommand{\PaperI}{\citetalias{Aird17}}
\defcitealias{Aird18}{Paper~II}
\newcommand{\PaperII}{\citetalias{Aird18}}
 
\newcommand{\upd}[1]{#1}\newcommand{\mupd}[1]{#1}\newcommand{\mmupd}[1]{#1}
\newcommand{\rone}[1]{{#1}}

\begin{document}

\title[The incidence of AGN as a function of SFR]{X-rays across the galaxy population - III. The incidence of AGN as a function of star formation rate}
\author[J. Aird et al.]{J. Aird$^{1}$\thanks{james.aird@leicester.ac.uk}, A. L. Coil$^2$ and A. Georgakakis$^{3,4}$\\
$^1$Department of Physics \& Astronomy, University of Leicester, University Road, Leicester LE1 7RJ, UK\\
$^2$Center for Astrophysics and Space Sciences (CASS), Department of Physics, University of California, San Diego, CA 92093, USA\\
$^3$Max Planck Institute f\"{u}r Extraterrestrische Physik, Giessenbachstrasse, 85748 Garching, Germany\\
$^4$IAASARS, National Observatory of Athens, GR-15236 Penteli, Greece}
\pagerange{\pageref{firstpage}--\pageref{lastpage}} \pubyear{2019}

\maketitle
\label{firstpage}
\begin{abstract}
\rone{
We map the co-eval growth of galaxies and their central supermassive black holes in detail by measuring the incidence of Active Galactic Nuclei (AGN) in galaxies as a function of star formation rate (SFR) and redshift (to $z\sim4$). We combine large galaxy samples with deep \textit{Chandra} X-ray imaging to measure the probability distribution of specific black hole accretion rates ($L_\mathrm{X}$ relative to stellar mass) and derive robust AGN fractions and average specific accretion rates. First, we consider galaxies \emph{along} the main sequence of star formation. We find a linear correlation between the average SFR and both the AGN fraction and average specific accretion rate across a wide range in stellar mass ($\mathcal{M}_*\sim10^{8.5-11.5}\mathcal{M}_\odot$) and to at least $z\sim2.5$, indicating that AGN in main-sequence galaxies are driven by the stochastic accretion of cold gas. We also consider quiescent galaxies and find significantly higher AGN fractions than predicted, given their low SFRs, indicating that AGN in quiescent galaxies are fuelled by additional mechanisms (e.g. stellar winds). Next, we bin galaxies according to their SFRs \emph{relative} to the main sequence. We find that the AGN fraction is significantly elevated for galaxies that are still star-forming but with SFRs below the main sequence, indicating further triggering mechanisms enhance AGN activity within these sub-main-sequence galaxies. We also find that the incidence of high-accretion-rate AGN is enhanced in starburst galaxies and evolves more mildly with redshift than within the rest of the galaxy population, suggesting mergers play a role in driving AGN activity in such high-SFR galaxies. 
}
\end{abstract}

\section{Introduction}
\label{sec:intro}

The growth of galaxies via star formation and the growth of the supermassive black holes that lie at their centres via periods of accretion that produce an Active Galactic Nucleus (AGN) appear to have proceeded in a broadly coherent manner over cosmic time, at least in a global sense. 
Both the total star formation rate (SFR) density and the total AGN accretion rate density peak at a redshift $z\sim1-3$ and decline at roughly the same rate towards more recent cosmic times \citep[e.g.][]{Boyle98,Madau14,Aird15}.
In addition, the mass of central black holes and the properties of their host galaxies (e.g. stellar velocity dispersion, bulge mass, total stellar mass) in the local universe are correlated \citep[e.g.][]{Ferrarese00,Gebhardt00,Kormendy13}, indicating that black hole and galaxy growth must have proceeded in a broadly consistent manner when averaged over the lifetime of a galaxy.

Identifying a connection between the \emph{instantaneous} level of AGN accretion and the current level of star formation---directly indicating that the growth of an individual galaxy and its central black hole are proceeding in unison---has proved much more difficult.
A number of studies focus on selecting AGNs (e.g. based on their X-ray emission), measuring the SFRs of their host galaxies, and exploring any correlation between the AGN luminosity and the SFR \citep[e.g.][]{Silverman09b,Page12,Rosario12}.
Such studies find that AGN of a given luminosity have a broad range of SFRs and are now starting to reach a consensus that there is very little, if any, correlation between the AGN luminosity and the SFR in such samples \citep[e.g.][]{Harrison12,Rosario13c,Azadi15}. 
Indeed, AGN with low-to-moderate luminosities are found to have average SFRs consistent with star-forming galaxies at the same redshift, once matched in stellar mass \citep[e.g.][]{Rosario13b,Stanley15,Stanley17}. 

These studies indicate there is little direct connection between star formation and black hole growth in individual galaxies.
\rone{However, such studies---adopting flux-limited samples of AGN---will select only those galaxies that currently contain an actively accreting black hole.
Recent observations have shown that galaxies with similar properties (e.g.~stellar masses, SFRs) contain AGN with a very broad range of accretion rates \citep[e.g.][]{Aird12,Bongiorno12,Azadi15}.
These broad distributions blur out any direct correlation between galaxy properties and the observed AGN luminosity in flux-limited AGN samples \citep[see][and discussion therein]{Stanley15,Bernhard18,Caplar18}.
The broad distributions may be due to variability of AGN activity on short timescales compared to galaxy-wide processes within individual galaxies \citep[i.e. $\lesssim$100~Myr;][]{Mullaney12b,Aird13,Hickox14}. 
Thus, all galaxies may host an AGN at some point during their lifetime but the AGN will ``flicker" on timescales that cannot be observed directly, making it difficult to determine how the overall properties of galaxies are connected to the growth of their central black holes. 
Such variability may reflect the stochastic nature of the processes that drive gas into the centres of galaxies to trigger AGN activity as well as the stability of the accretion onto the black hole itself \citep[e.g.][]{Hopkins06c,Novak11,King15,Yuan18}, although the exact timescales and evolution of AGN episodes in individual galaxies remain poorly constrained by observations.
Indeed, the observed distributions could be consistent with long-lived AGN activity at different fixed accretion rates in a subset of galaxies although such a model is difficult to reconcile with the ubiquity of supermassive black holes throughout the galaxy population. 
Regardless, the broad distribution of accretion rates must be accounted for to reveal any underlying connections between the SFRs of galaxies and their overall levels of AGN activity.}

A number of different approaches have been adopted to account for \rone{the broad distribution of AGN accretion rates.}
Large samples of galaxies with similar properties (e.g. SFRs, stellar masses) can be constructed and a stacking analysis can be used to measure the \emph{average} AGN luminosity, which reveals a correlation between the SFRs of galaxies and their black hole growth \citep[e.g.][]{Mullaney12b,Chen13,Rodighiero15,Yang17}. 
Alternatively, a modelling approach can be adopted to connect the global properties of the galaxy population (e.g. described by their stellar mass function) and the observed properties of AGN samples, inferring an underlying distribution of black hole accretion rates \citep[e.g.][]{Aird13,Veale14,Caplar15,Caplar18,Bongiorno16,Georgakakis17,Bernhard18}. 
We have pioneered a third approach -- taking large samples of galaxies with similar properties and directly measuring the distribution of accretion rates within such galaxies using deep X-ray data (e.g.~\citealt{Aird12,Azadi15,Aird18}; see also \citealt{Bongiorno12,Yang18}). 
Our approach places direct observational constraints on the fraction of galaxies that contain an AGN with different accretion rates, tracing the incidence of AGN and their varying rates of accretion across the galaxy population.

In this paper, we build on this approach to investigate how the incidence of AGN is related to the SFRs of galaxies. 
This paper forms the third in a series exploring the X-ray emission from near-infrared selected galaxy samples in the premier extragalactic survey fields.
\PaperI\ \citep{Aird17} presented the intrinsic distribution of X-ray luminosities for samples of star-forming galaxies as a function of stellar mass and redshift, identifying peaks in these distributions at low luminosities ($\lx\lesssim10^{42}$~\ergs) that were used to trace the SFR based on the X-ray emission.
In \PaperII\ \citep{Aird18} we used these same datasets to track the levels of AGN activity within galaxy samples---traced by X-ray emission at higher luminosities---and measured the distribution of specific black hole accretion rates (hereafter, \Sar) as a function of stellar mass (from $10^{8.5}$ to $10^{11.5} \msun$) and redshift (to $z\sim4$) for both star-forming and quiescent galaxy samples. 
These new measurements identify a broad distribution of \Sar\ in both galaxy types (and across the full range of stellar mass and redshift down to dwarf galaxies at $z<1$ and massive galaxies to $z\sim4$), likely reflecting the variability of AGN activity. 
Our measurements also reveal dependences on stellar mass and redshift that differ between star-forming galaxies and quiescent galaxies.
Specifically, the probability of a star-forming galaxy hosting a moderate accretion rate AGN increases with stellar mass and evolves strongly with redshift (for moderate-to-high stellar masses). 
Conversely, the probability of a quiescent galaxy hosting an AGN shows signs of suppression at the highest stellar masses and is generally lower than in star-forming galaxies of equivalent stellar mass and redshift.
These differences suggest that different physical mechanisms may drive the fuelling of AGN activity in the different galaxy populations (e.g. stochastic accretion of cold gas in star-forming galaxies versus stellar winds in quiescent galaxies).

This paper builds on the analysis of \PaperII\ to explore in greater detail the connection between AGN activity and the SFRs of their host galaxies.
Section~\ref{sec:data} gives an overview of our datasets
and Section~\ref{sec:method} gives an overview of our methodology, first presented in \PaperII, to combine the available X-ray data for all galaxies in a sample, measure the probability distributions of \Sar, \mupd{and derive robust AGN fractions and average specific accretion rates, characterising the incidence of AGN across the galaxy population.}
\upd{In Section~\ref{sec:fagn_sfandqu} we investigate how the incidence of AGN in star-forming galaxies changes as SFRs increase \emph{along} the main sequence of star formation; we also compare to the incidence of AGN in quiescent galaxies of different stellar masses and SFRs.
In Section~\ref{sec:fagn_vs_mainseq} we investigate how the incidence of AGN changes \emph{across} and below the main sequence,
dividing our galaxy samples into five populations according to their SFRs relative to the main sequence.}
We discuss our findings in Section~\ref{sec:discuss} and summarize in Section~\ref{sec:summary}.
Throughout this paper, we assume a \citet{Chabrier03} stellar initial mass function when deriving galaxy properties and adopt a flat cosmology with $\Omega_\Lambda = 0.7$ and $H_0 = 70$~km~s$^{-1}$~Mpc$^{-1}$.

\section{Data and sample selection}
\label{sec:data}

The results in this paper are based on the same datasets used in \PaperI\ and \PaperII.
We provide a brief summary here (see \PaperI\ for full details).

We adopt samples of galaxies selected using deep near-infrared (NIR) imaging of four of the CANDELS fields \citep[GOODS-S, GOODS-N, AEGIS and COSMOS:][]{Koekemoer11,Grogin11} as well as the larger area UltraVISTA survey \citep{McCracken12}. 
We adopt the NIR-selected catalogues provided by \citet{Skelton14} and \citet{Muzzin13} for the CANDELS and UltraVISTA fields, respectively, which provide aperture-matched photometry spanning from the UV to mid-infrared (MIR). 
We combine these data with deep \textit{Chandra} X-ray imaging 
which have been reprocessed using our own procedures \citep[see][\PaperI]{Laird09,Nandra15,Aird15}. 
We cross-match the NIR-selected catalogues with significant X-ray detections in the 2--7~keV band and extract data (total counts, background counts, and effective exposures) from the same energy band at the positions of the remaining NIR-selected sources. 
The X-ray data from both detections and non-detections are combined using a Bayesian methodology that allows us to probe below the nominal sensitivity limits, correct for incompleteness, and recover the intrinsic distribution of accretion rates within a given sample of galaxies (see Section~\ref{sec:method} below and \PaperII\ for full details).   
We compile spectroscopic and photometric redshifts for the NIR-selected sources, as described in \PaperI, including high quality AGN photo-$z$ for X-ray detected sources \citep[see][]{Hsu14,Nandra15,Marchesi16}. 

To derive the physical properties of galaxies (stellar masses, SFRs), we fit the observed UV-to-MIR spectral energy distributions using our updated version of the FAST code,\footnote{FAST was originally developed by \citet{Kriek09}. Our updated version, which allows for both galaxy and AGN components, is described in the appendices of \PaperI\ and \PaperII\ and is available at \url{https://github.com/jamesaird/FAST}.}
allowing for an AGN component for sources with significant X-ray detections. 
We estimate SFRs based on either the sum of the UV and IR emission (for sources with 24\micron\ detections) or the SFR recovered from the SED fitting (for sources without 24\micron\ detections), in both cases accounting for any AGN contamination (see appendix~A of \PaperII).
We exclude sources from our sample that are identified as AGN-dominated based on the SED fitting (where $>50$ per cent of the light at rest-frame 5000\AA\ is associated with the AGN component) as it is not possible to recover an accurate estimate of the SFR. 
Such sources constitute $\lesssim$20 per cent of the X-ray detected sample and $\lesssim$0.3 per cent of the overall galaxy sample.
We do not expect the exclusion of these AGN-dominated sources to have a significant impact on our results or overall conclusions (see also appendix~C of \PaperII).

We define our sample of galaxies as in \PaperII, excluding stars and applying the NIR magnitude limits and stellar mass completeness limits described in detail in \PaperI. 
\rone{The stellar mass completeness limits are defined such that a maximally red galaxy with an extremely low SFR would still be detected in our NIR imaging and thus enter our sample, ensuring we have complete samples of quiescent galaxies.}
In Section~\ref{sec:fagn_sfandqu} we use the full sample of 126,971 galaxies, including both star-forming and quiescent galaxies and spanning a  broad range in stellar mass ($8.5<\log \mstel/\msun<11.5$) and redshift ($0.1< z < 4$). 
1797 of these galaxies are directly detected in the 2--7~keV band in our deep \textit{Chandra} imaging. 
In Section~\ref{sec:fagn_vs_mainseq} we further restrict our analysis to a single stellar mass bin ($10.0<\log \mstel/\msun < 11.5$) and divide our galaxies according to their SFRs relative to the main sequence of star formation (see Section~\ref{sec:fagn_vs_mainseq} below for full details). 
This stellar mass bin lies wholly above the completeness limits of our deep CANDELS NIR imaging out to $z\gtrsim4$, ensuring we can track changes in the levels of AGN accretion as a function of SFR at a fixed stellar mass. 
The UltraVISTA galaxy sample is incomplete for galaxies with $\mstel>10^{10}\msun$ at $z\gtrsim2$; only galaxies lying above the redshift-dependent stellar mass completeness limit of UltraVISTA are included in our sample, supplemented by lower mass galaxies from the deeper CANDELS imaging. 
We note that galaxies containing X-ray AGN are subjected to the same mass completeness limits, ensuring our measurements accurately reflect the AGN content of the galaxy sample.

\section{Methodology: accretion rate probability distributions, \mupd{AGN fractions and average specific accretion rates}}
\label{sec:method}

In this work (as in \PaperII), we describe the AGN content of our galaxy samples using measurements of $p(\log \sar\giv \mstel, z, \mathrm{SFR})$, the probability density function of specific black hole accretion rates, \Sar, within a sample of galaxies of given stellar mass, redshift, and SFR. 
We define this function per unit dex and it is normalized such that
\begin{equation}
\int p(\log \sar\giv \mstel, z, \mathrm{SFR})\; d \log \sar = 1
\label{eq:psar}
\end{equation}
and is hereafter referred to more succinctly as \Psar. 
The probability density function, \Psar, characterizes both the overall distribution of \Sar\ within a given sample of galaxies \emph{and} the fraction of galaxies that host AGN of different \Sar, enabling a detailed comparison of AGN activity in different galaxy populations.

We choose to present our measurements in terms of \Sar, the black hole accretion rate as traced by the X-ray luminosity normalized relative to the stellar mass of the host galaxy, as defined by \citet[see also \PaperII, \citealt{Bongiorno16}, \citealt{Georgakakis17}]{Aird12} and given by 
\begin{equation}
\sar  = \frac{ k_\mathrm{bol} \; L_\mathrm{X} }
                  { 1.3 \times 10^{38} \; \mathrm{erg\;s^{-1}} \times 0.002 \dfrac{\mathcal{M}_*}{\mathcal{M}_\odot} }
\label{eq:sar}
\end{equation}
where $L_\mathrm{X}$ is the rest-frame 2--10~keV X-ray luminosity (originating from the central AGN), $k_\mathrm{bol}$ is a bolometric correction factor (as in \PaperII, we adopt a constant $k_\mathrm{bol}=25$) and $\mathcal{M}_*$ is the stellar mass of the host galaxy. 
By measuring the \emph{specific} accretion rate, normalized by the stellar mass, we are able to account for the selection bias whereby a black hole growing at a given accretion rate in a low mass galaxy produces a lower observable X-ray luminosity (and is thus harder to detect) than a black hole in a higher mass galaxy growing at the same specific accretion rate \citep{Aird12}.
The additional factors in Equation \ref{eq:sar} are chosen such that $\sar\approx\lambda_\mathrm{Edd}$, the Eddington ratio of the black hole, under the assumption that the black hole mass ($\mathcal{M}_\mathrm{BH}$) is directly proportional to the galaxy stellar mass, $\mathcal{M}_\mathrm{BH} = 0.002 \mstel/\msun$.
We acknowledge that uncertainties, scatter, and systematic changes in this scaling would alter the relationship between \Sar\ and $\lambda_\mathrm{Edd}$.
Nonetheless, by measuring \Sar\ we retain an observable quantity (effectively $\sar\propto \lx/\mstel$) that reflects the rate that the central black hole is growing in a given galaxy. 

\begin{figure}
\begin{center}
\includegraphics[width=0.88\columnwidth,trim=30 40 30 30]{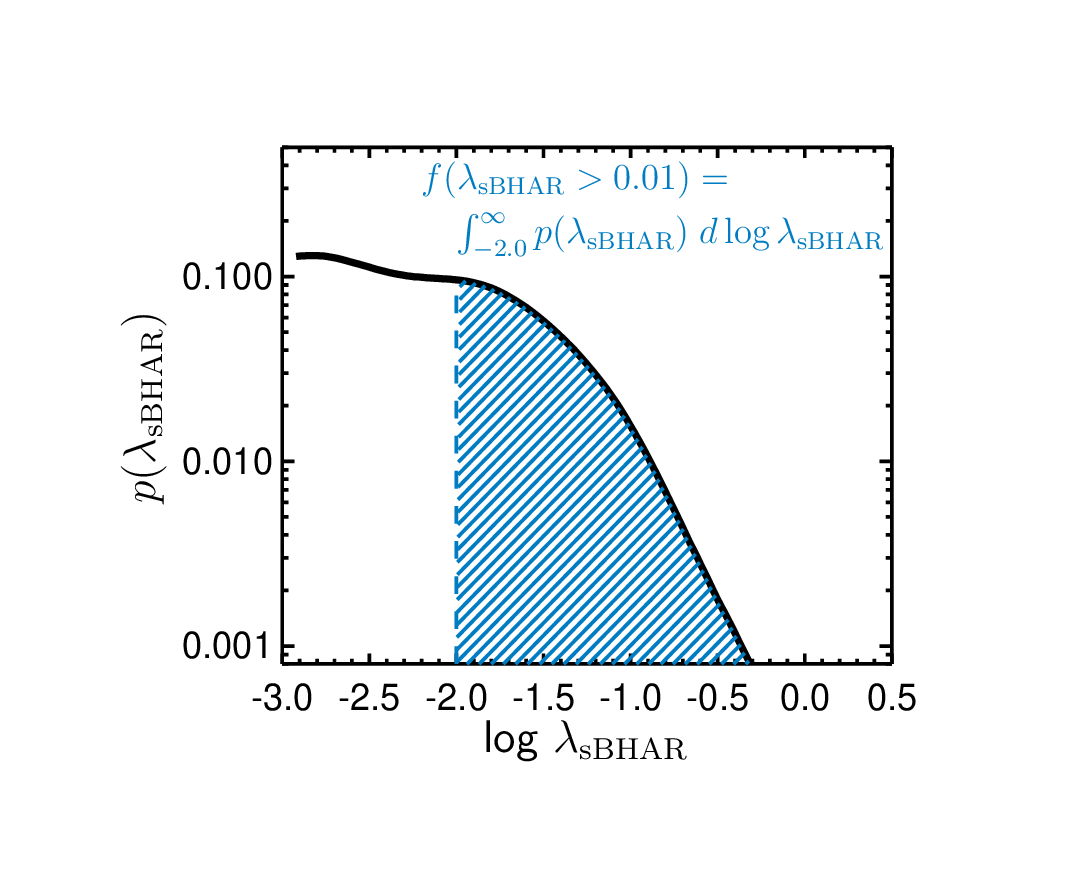}
\includegraphics[width=0.88\columnwidth,trim=30 40 30 30]{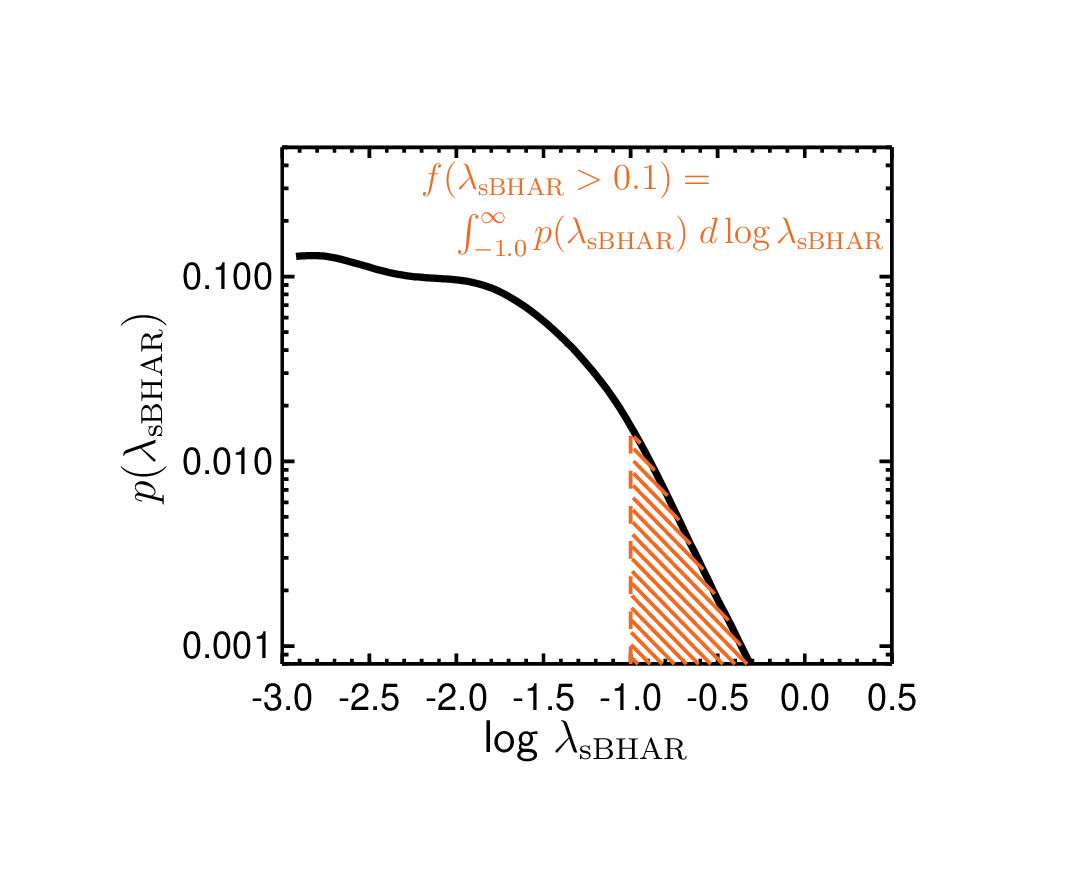}
\includegraphics[width=0.88\columnwidth,trim=30 40 30 30]{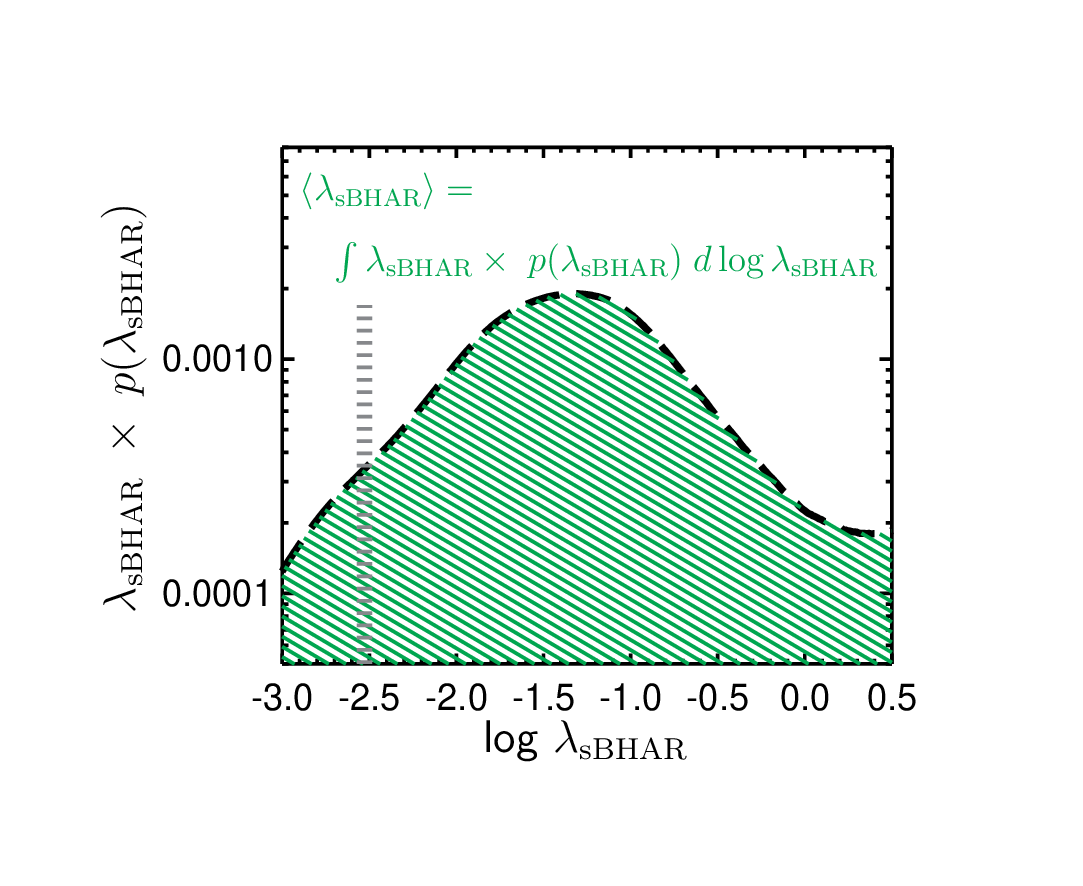}
\end{center}
\caption{
\mupd{
Illustrations showing how we summarize our measurements of \Psar\ by calculating three quantities to describe the ``incidence of AGN."
\emph{Top panel}: AGN fraction, \Fduty, which is calculated by integrating \Psar\ (solid black line) down to a limit of $\sar=0.01$.
\emph{Middle panel}: High-\Sar AGN fraction, \Fbright, calculated by integrating \Psar\  (solid black line) to a limit of $\sar=0.1$ and thus indicating the smaller fraction of galaxies that contain the most rapidly accreting AGN.
\emph{Bottom panel}: 
The average specific black hole accretion rate, \Avsar, found by multiplying \Psar\ by \Sar\ (shown by the dashed black line) and integrating. 
\rone{The average of the distribution is indicated by the vertical dotted grey line. The measurement is dominated by AGN around the break in \Psar\ but also accounts for the large fraction of the probability distribution that corresponds to periods of very low \Sar\ (below $\log\sar=-3$) and thus is not aligned with the peak of $\sar\times p(\sar)$.}
}}
\label{fig:method}
\end{figure}

We use deep 2--7~keV \textit{Chandra} X-ray data to measure \Psar,
combining X-ray detections and X-ray information at the positions of the remaining galaxies that lack direct detections using a Bayesian methodology (described in detail in the appendices to \PaperI\ and \PaperII). 
Our method is non-parametric, describing \Psar\ using a flexible combination of 36 evenly spaced component functions (specifically, gamma functions that are conjugate to the Poisson likelihood that describes the observed X-ray data).
We require that integrating over all components (extending down to $\log \sar=-7$) equates to 1 (satisfying Equation~\ref{eq:psar} above) and adopt a prior that prefers a smoothly varying distribution to link the different components. 
This approach is highly flexible and does not impose a specific functional form, ensuring that the data determine the overall shape of the recovered distribution.
Our statistical method accounts for the uncertainties in the X-ray fluxes of individual sources and---by including X-ray information from non-detections---allows us to probe below the nominal sensitivity limits and include data from lower significance sources that would not enter a traditional X-ray selected catalogue. 
Most crucially, our method corrects for the varying sensitivity of the X-ray imaging, both between different fields and across the area of a single field. 
We apply an additional correction to account for the X-ray emission associated with stellar processes (primarily, high- and low-mass X-ray binaries) as described in appendix~B of \PaperII. 
However, as in \PaperII, we only present measurements of \Psar\ that probe $>0.5$~dex above the ``X-ray main sequence of star formation" identified in \PaperI\ ($\lx\gtrsim10^{41-42}$~\ergs, depending on stellar mass and redshift), ensuring we are probing the distribution of X-ray emission related to AGN activity.

\mupd{
In this paper, we derive three parameters to describe the ``incidence of AGN" within a given galaxy sample and summarize our full measurements of \Psar: the AGN fraction to two different limits in \Sar\ and the average specific accretion rate, \Avsar.
These parameters summarize the full \Psar\ distributions and allow us to investigate trends with galaxy parameters such as stellar mass or redshift.
Figure~\ref{fig:method} provides a visual illustration of how these parameters are calculated from our measurements of \Psar\ for a given galaxy sample. 
}

We first calculate robust ``AGN fractions", which describe the fraction of a given sample of galaxies that contain a central black hole that is accreting above a fixed limit in \Sar.\footnote{\upd{These AGN fractions may also be interpreted as a ``duty cycle", describing the fraction of its lifetime that a certain type of galaxy hosts an AGN, and were referred to as such in \PaperII.}}
We calculate the AGN fraction by integrating our measurements of \Psar. Thus, our AGN fraction is given by
\begin{equation} 
f(\sar > \lambda_\mathrm{lim}) = \int^\infty_{\log \lambda_\mathrm{lim}} p(\sar) \;\mathrm{d}\log \sar
\end{equation}
where $\lambda_\mathrm{lim}$ indicates the adopted limit on \Sar.
In this paper, we present measurements of the AGN fraction to two different limits:
\begin{enumerate}
\item
$f(\sar> 0.01)$, roughly corresponding to black holes growing above $\sim$1~per~cent of their Eddington limit (assuming the nominal scalings given in Equation \ref{eq:sar}) and thus indicating the fraction of galaxies with a moderate-to-high accretion rate, X-ray selected AGN (hereafter referred to as the generic ``AGN fraction"); and 
\item 
$f(\sar > 0.1)$, corresponding to the fraction of galaxies with a high-accretion-rate AGN, growing their black holes at $\gtrsim$~10~per~cent of their Eddington limit and hereafter referred to as the ``high-\Sar\ AGN fraction".
\end{enumerate}
These measurements of AGN fractions, based on our measurements of \Psar, fully account for the varying sensitivity of the X-ray observations across the CANDELS and UltraVISTA fields and are thus robust and well-defined compared to simple detection fractions \citep[that are highly dependent on the depths of the available data, see e.g.][]{Mendez13}.
Furthermore, by defining AGN fractions to fixed limits in \Sar, rather than fixed limits in X-ray luminosity, we account for the stellar-mass-dependent selection bias identified by \citet{Aird12} and discussed above, ensuring our measurements can be meaningfully compared across a wide range in stellar mass.

\upd{
The average specific black hole accretion rate within a given galaxy sample, \Avsar, (referred to hereafter as the ``average specific accretion rate") is derived from our measurements of \Psar\ by}
\begin{equation}
\avsar  = \int_{-\infty}^{\infty} \sar \;p(\sar) \; d \log \sar.
\end{equation}
\mupd{
This measurement combines all the X-ray data for a given galaxy sample that is used in our measurements of \Psar.
As such, these measurements are similar to performing a stacking analysis but use the information from both detections and non-detections in a consistent manner.}

\mupd{
Our measurement of \Avsar\ provides a tracer of the typical rate of black hole growth in a given galaxy sample. 
However, we know that AGN activity is a transient phenomenon and that significant AGN activity is usually only observed in a small fraction of galaxies at a given time.
Our measurements of AGN fractions, combined with \Avsar, thus provide a comprehensive description of the incidence of AGN activity across the galaxy population, summarizing our detailed measurements of \Psar.
We estimate 1$\sigma$-equivalent (i.e. 68 per cent) confidence intervals on \Fduty, \Fbright\ and \Avsar\ by propagating the uncertainties in \Psar\ given by our Bayesian analysis.}

\section{The incidence of AGN as a function of average star formation rate in star-forming and quiescent galaxies}
\label{sec:fagn_sfandqu}

\upd{
In this section, \mupd{we measure AGN fractions and average specific accretion rates in samples of star-forming galaxies}, divided according to their position \emph{along} the main sequence of star formation and thus allowing us to probe the dependence on both stellar mass and SFR (which are correlated for star-forming galaxies on the main sequence).
We compare these results to the incidence of AGN in 
quiescent galaxies over the same range of stellar mass and redshift, which, by definition, have substantially lower SFRs.  
We use our previous measurements of \Psar\ for both star-forming and quiescent galaxies as a function of stellar mass and redshift, first presented in \PaperII, to derive the AGN fractions and average specific accretion rates presented here.
In Section~\ref{sec:fagn_vs_mainseq} below, we present new measurements of \Psar, average specific accretion rates and AGN fractions, investigating \emph{across} and below the main sequence by dividing galaxies at a fixed stellar mass into five samples according to their SFRs \emph{relative} to the main sequence. 
}

\begin{figure}
\includegraphics[width=\columnwidth,trim=40 30 50 10]{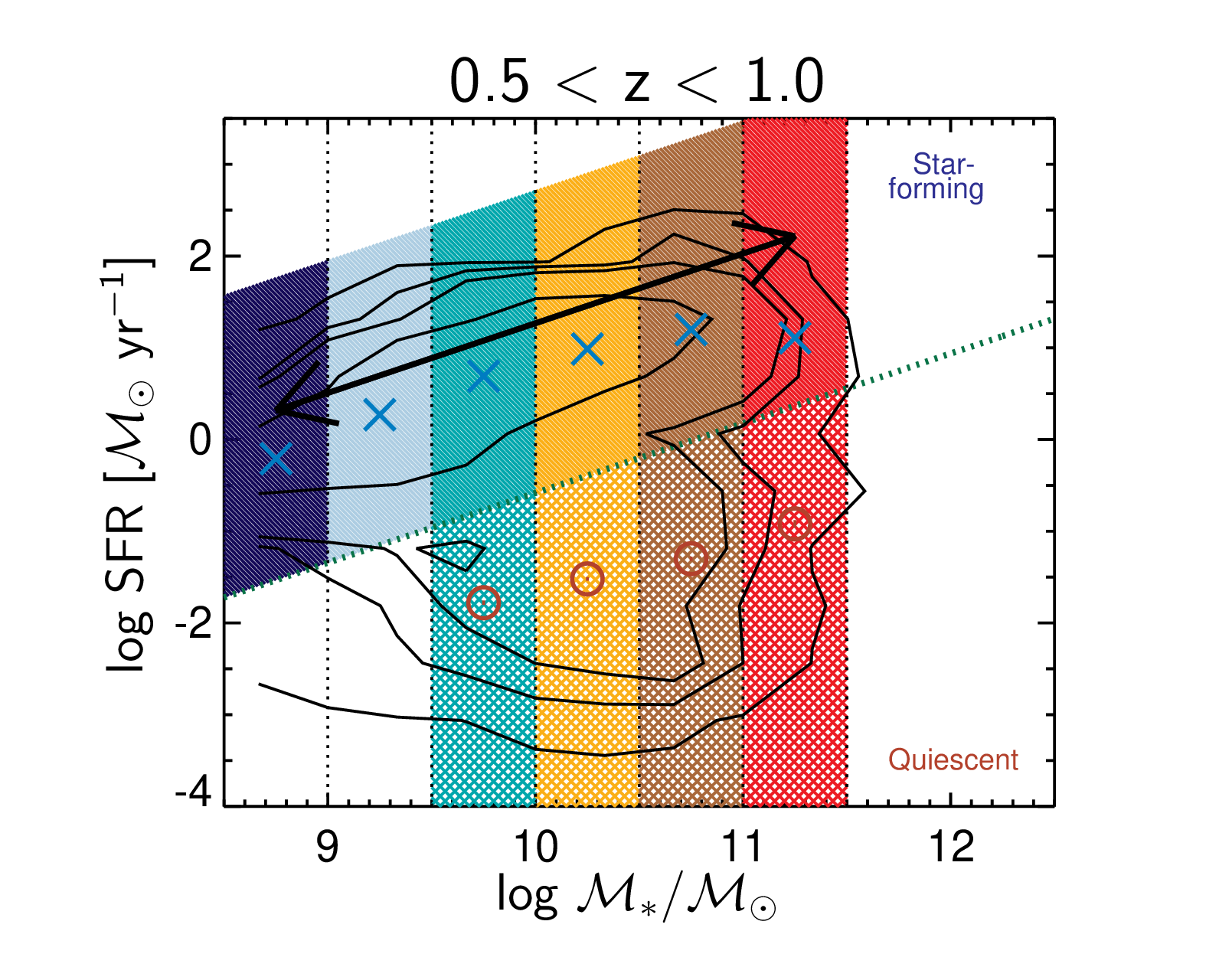}
\caption{
The distribution of SFRs as a function of \Mstel\ for our galaxy sample (black contours which enclose 68, 90, 95 and 99~per~cent of galaxies) for one of our redshift bins, $0.5<z<1.0$. The green dotted line, set at 1.3 dex below the evolving star-forming main sequence, separates the star-forming and quiescent galaxy populations, as indicated. 
The coloured regions indicate our stellar-mass-dependent binning of each galaxy population. 
Blue crosses indicate $\langle \log \mathrm{SFR}\rangle$ of the star-forming galaxies in each mass bin and show how the increasing stellar mass bins, \upd{moving \emph{along} the main sequence}, contain galaxies with progressively higher average SFRs. 
\upd{Red circles show the average SFRs of the quiescent galaxies in each stellar mass bin, which---by definition---are lower than the SFRs of the star-forming galaxies of equivalent mass and also span a range of values across the different stellar masses.}
Our measurements of the AGN fraction and average specific accretion rate in each stellar-mass and redshift bin, based on the measurements of \Psar\ from \PaperII, thus allow us to explore the incidence of AGN as a function of \upd{both stellar mass and SFR} (see Figures~\ref{fig:fagn_vs_mstel}--\ref{fig:fduty_vs_sfr_onepanel_sfqu}).
}  
\label{fig:sfr_vs_mstel_bins1}
\end{figure}

\upd{
Figure~\ref{fig:sfr_vs_mstel_bins1} illustrates the binning of the galaxy population in one of our redshift ranges adopted in this section.
We divide our galaxy sample into star-forming and quiescent galaxies using a cut in SFR that corresponds to 1.3~dex below the stellar-mass- and redshift-dependent main sequence of star formation (see equation 1 of \PaperII).
Our galaxy sample is further divided into bins of stellar mass; we thus probe increasing SFRs as we move along the main sequence of star formation.
}
The blue crosses in Figure~\ref{fig:sfr_vs_mstel_bins1} indicate the geometric mean SFR, $\langle \log \mathrm{SFR} \rangle$ or hereafter the ``average SFR", for the star-forming galaxies in each stellar mass bin.\footnote{We note that our measurements of the average SFRs of star-forming galaxies are consistent with a mild flattening in the slope of the main sequence of star formation for $\log \mstel/\msun\gtrsim10$ at $z\lesssim1$ \citep[see e.g.][]{Whitaker14,Schreiber15,Lee15}; thus our highest stellar mass bins at $z<1$ tend to have similar average SFRs.} By definition, the quiescent galaxy samples all have lower SFRs 
\upd{yet still span a range of $\sim$1~dex in SFR (at a given redshift) across the different stellar mass bins.}
In our higher redshift bins, the star-forming galaxies shift to progressively higher SFRs due to the evolution of the main sequence; 
the quiescent galaxy samples also have higher average SFRs at higher redshifts. 

\begin{figure}
\begin{center}
\includegraphics[width=0.95\columnwidth,trim=10 50 40 40]{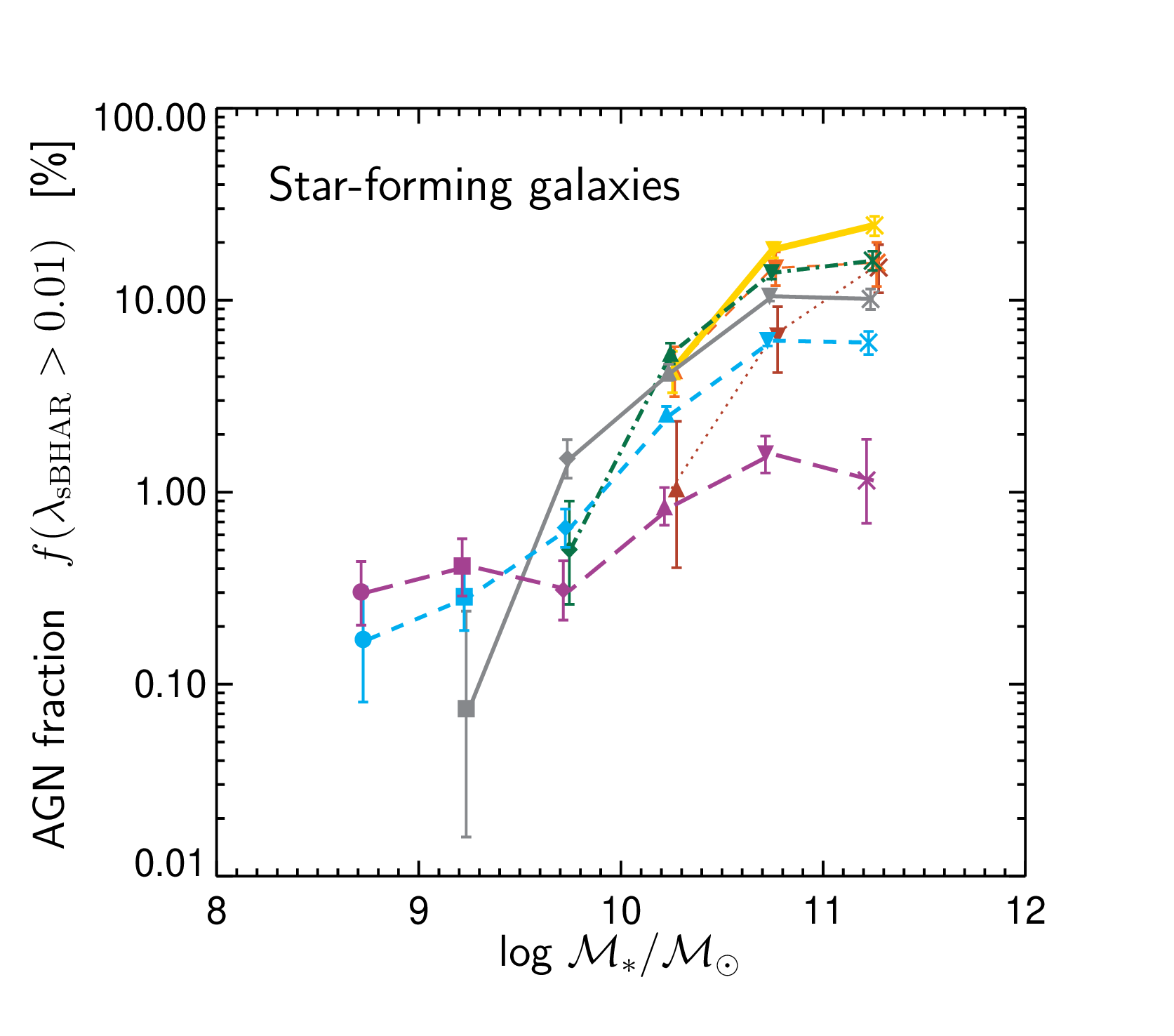}
\includegraphics[width=0.95\columnwidth,trim=10 50 40 10]{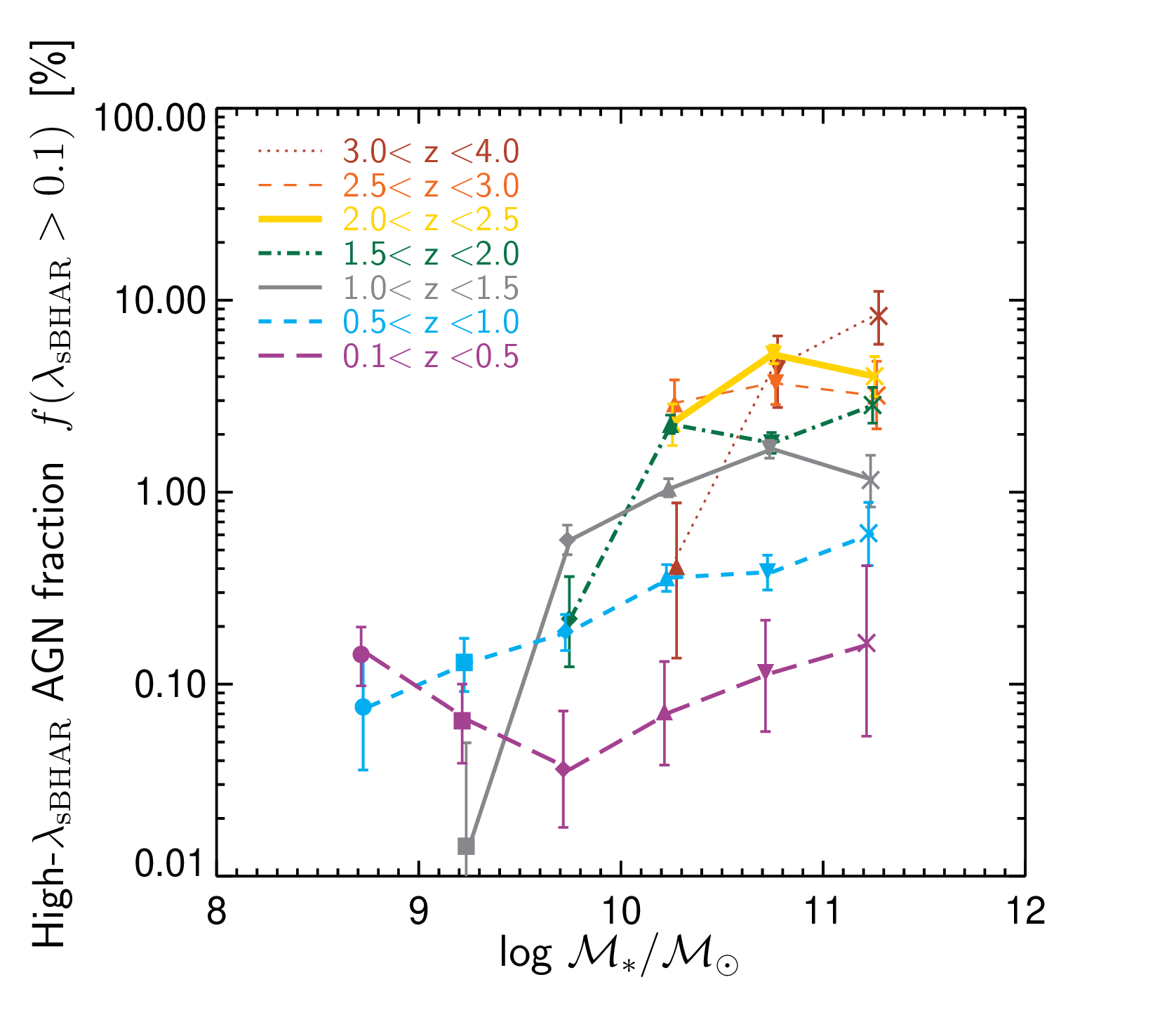}
\includegraphics[width=0.95\columnwidth,trim=10 40 40 10]{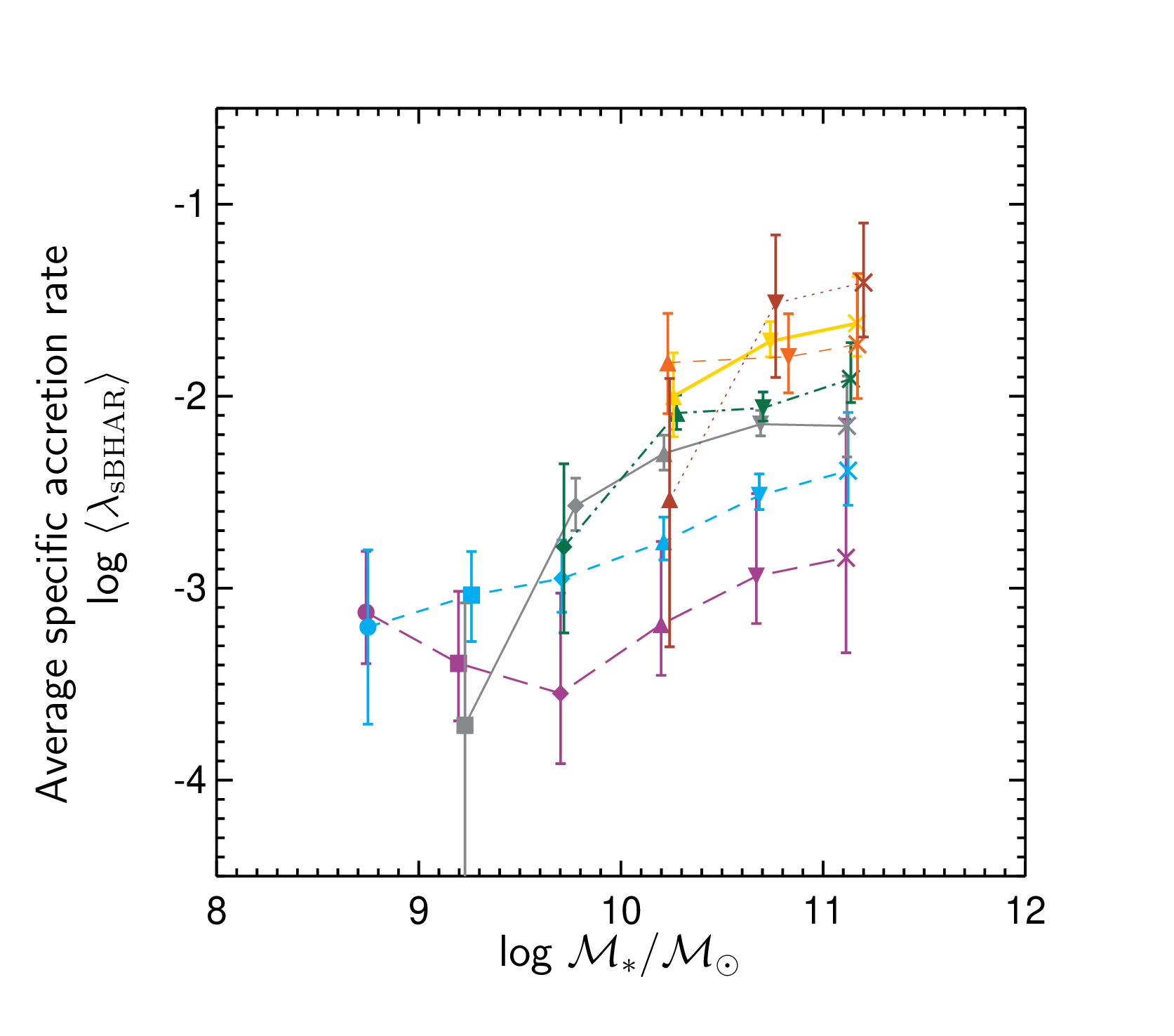}
\end{center}
\caption{
\mupd{
Measurements of the AGN fraction, \Fduty\ [\emph{top panel}], high-\Sar\ AGN fraction, \Fbright\ [\emph{middle panel}], and average specific accretion rate, \Avsar [\emph{bottom panel}], in star-forming galaxies as a function of stellar mass at different redshifts (as indicated by the colours).
These measurements are based on the estimates of \Psar\ previously presented in \PaperII.
The AGN fractions and average specific accretion rates generally increase toward higher stellar mass (at fixed $z$). 
At fixed stellar mass, we find a strong evolution with redshift (at least for $\log \mstel/\msun\gtrsim10$).
The AGN fraction increases between $z\sim0.1$ and $z\sim2$ and may decline toward higher redshift, whereas the 
average specific accretion rate continues to evolve to higher values at higher~$z$.
}}
\label{fig:fagn_vs_mstel}
\end{figure}
\begin{figure}
\begin{center}
\includegraphics[width=0.95\columnwidth,trim=10 50 40 40]{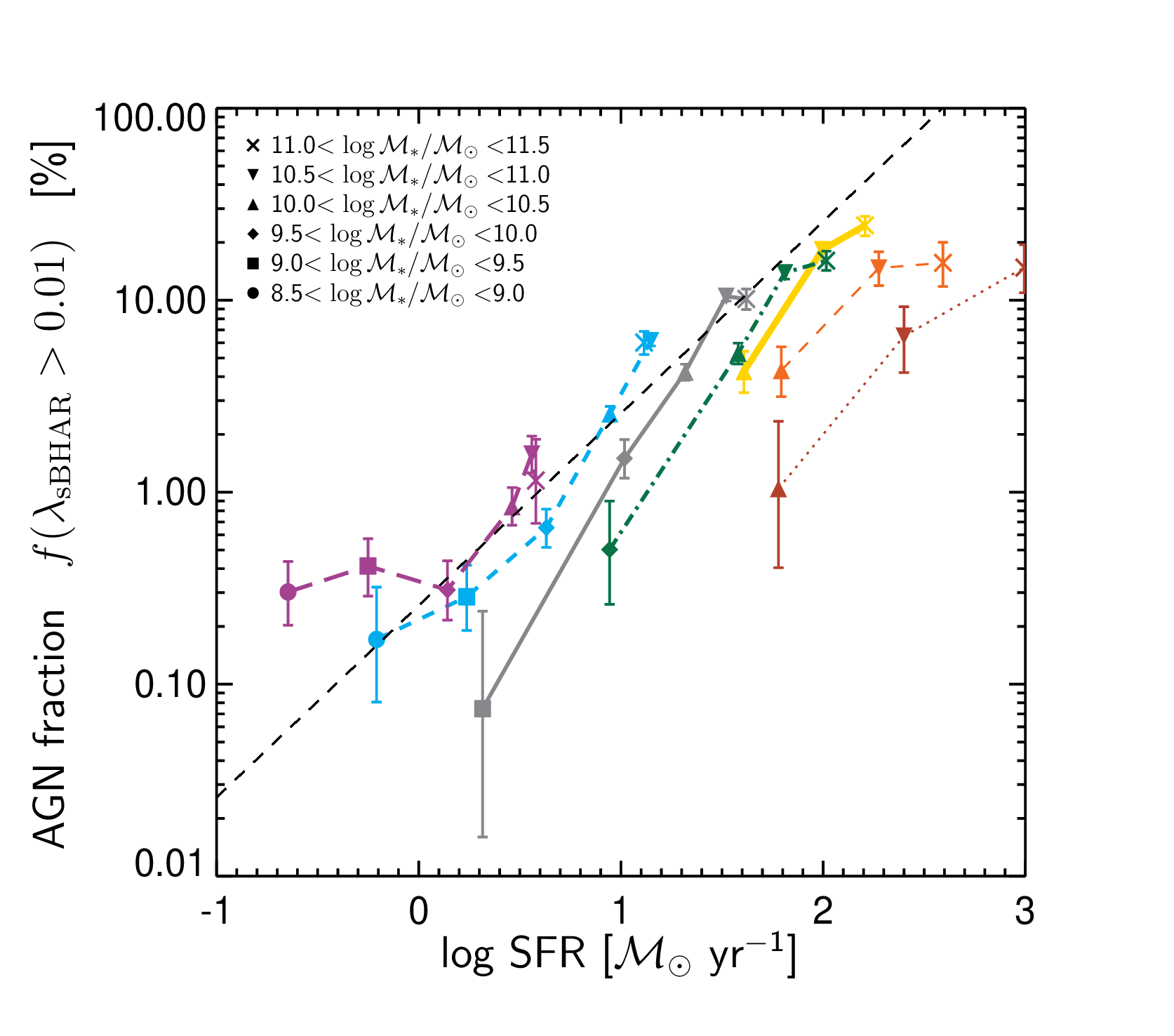}
\includegraphics[width=0.95\columnwidth,trim=10 50 40 10]{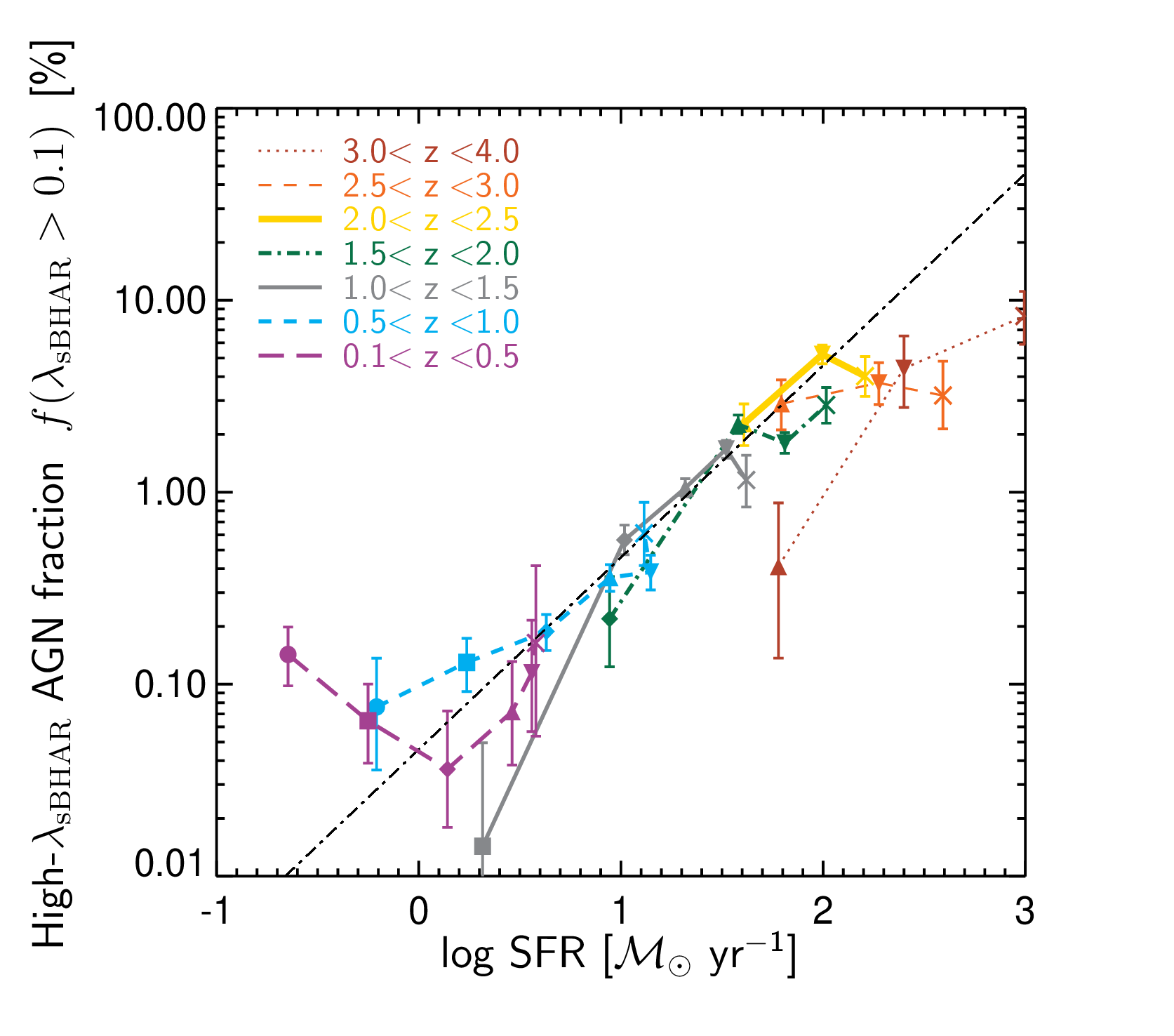}
\includegraphics[width=0.95\columnwidth,trim=10 40 40 10]{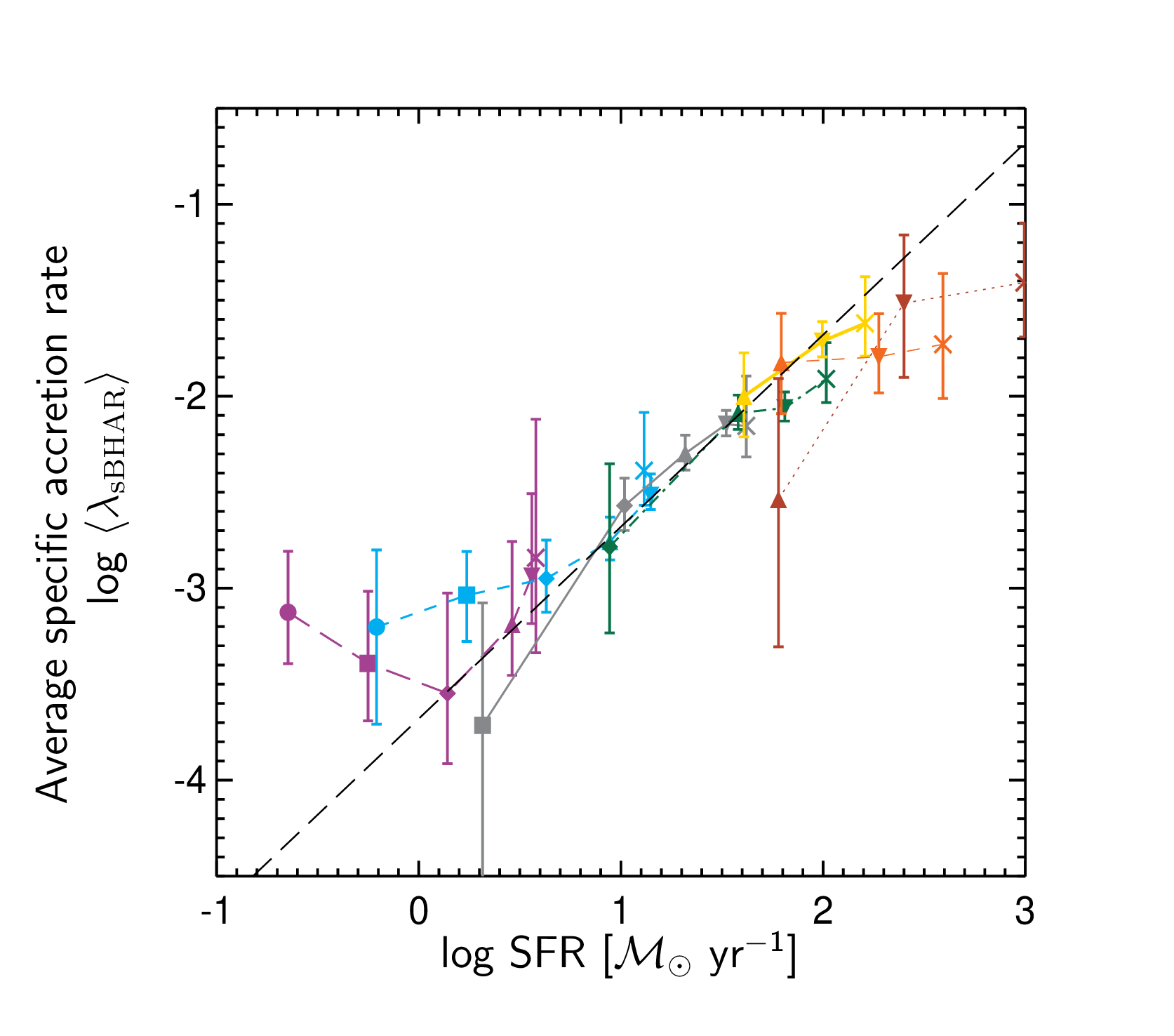}
\end{center}
\caption{
\upd{
Measurements of the AGN fraction (\emph{top}), high-\Sar\ AGN fraction (\emph{middle}), and average specific accretion rate (\emph{bottom}) in star-forming galaxies, as in Figure~\ref{fig:fagn_vs_mstel} but now shown as a function of the average SFR of each star-forming galaxy sample.
We find an approximate 1:1 correlation between all three quantities and the SFR out to at least $z\sim2$, as shown by the dashed, dot-dashed and long-dashed lines in the top, middle and bottom panels respectively.
Both the stellar mass dependence of the AGN fractions and the redshift evolution (to $z\sim2$) can be broadly explained by the increase in the average SFR moving both \emph{along} the main sequence (i.e. to higher stellar mass and SFR at fixed $z$) and to higher $z$ (resulting in a higher SFR at fixed stellar mass). 
}}
\label{fig:fagn_vs_sfr}
\end{figure}

\upd{
Our measurements of the AGN fraction, \Fduty, the high-\Sar\ AGN fraction, \Fbright, and \mupd{the average specific accretion rate, \Avsar,} in star-forming galaxies are presented as a function of \Mstel\ in Figure~\ref{fig:fagn_vs_mstel} and as a function of average SFR in Figure~\ref{fig:fagn_vs_sfr}, summarizing the full measurements of \Psar\ that were presented in \PaperII.
The top panel of Figure~\ref{fig:fagn_vs_mstel} recalls our main findings from \PaperII, showing the AGN fraction as a function of \Mstel.
At a fixed redshift, we find that the AGN fraction increases toward higher stellar masses.
At fixed stellar mass (for $\mstel\gtrsim10^{10}\msun$), we find that the AGN fraction increases rapidly with redshift out to $z\sim2$ and may drop slightly toward higher redshift.
This redshift evolution appears somewhat stronger at higher stellar masses with little or no evolution found in the low-mass, dwarf galaxy regime~($\mstel\lesssim 10^{9.5}\msun$).
The middle panel of Figure~\ref{fig:fagn_vs_mstel} presents new estimates of the high-\Sar\ AGN fraction in star-forming galaxies as a function of \Mstel, based on the measurements of \Psar\ from \PaperII. 
The fraction of galaxies with a high-\Sar AGN, \Fbright, is naturally lower than \Fduty\ at the same stellar mass and redshift and the uncertainties in the measurements are slightly larger. 
Nonetheless, we find a similar stellar-mass and redshift dependence for the high-\Sar\ AGN fraction as seen in the measurements of \Fduty.
}
\mupd{
The bottom panel of Figure~\ref{fig:fagn_vs_mstel} presents measurements of \Avsar\ as a function of \Mstel. 
Similar to the AGN fractions, we find that the average specific accretion rate increases mildly with increasing stellar mass, similar to the high-\Sar\ AGN fraction and indicating such sources---representing rare periods of rapid accretion over the lifetime of a galaxy---dominate the overall rate of accretion. 
In addition, we find a significant increase in \Avsar\ with redshift that---in contrast to the AGN fractions---may continue to increase above $z\sim2$.
}

\upd{
In Figure~\ref{fig:fagn_vs_sfr} we present the same measurements for star-forming galaxies but as a function of average SFR. We note a number of important results based on these measurements.
Firstly, we find that the AGN fraction, the high-\Sar\ AGN fraction and the average specific accretion rate in star-forming galaxies 
all tend to follow a roughly linear correlation with the average SFR, at least out to $z=2.5$, in contrast to the more complex behaviour seen as a function of \Mstel\ and $z$ in Figure~\ref{fig:fagn_vs_mstel}.
Star-forming galaxies with higher SFRs---either due to the higher redshift (and the evolution of the main sequence) or a higher stellar mass (and the slope of main sequence)---tend to contain a higher fraction of AGN and thus have higher average specific accretion rates.
\mupd{
We note that the trend in the AGN fraction appears to saturate at the highest redshifts, where the correspondingly higher SFRs of the galaxies do \emph{not} result in an extremely high AGN fraction. 
This effect is seen most clearly for \Fduty; the high-\Sar\ AGN fraction and average specific accretion rate do not appear to be affected and are well correlated with the average SFR over the range of redshifts that we probe.}
}
\upd{Assuming a basic one-to-one correlation, we perform a linear $\chi^2$ to our measurements as a function of SFR to $z=2.5$.
We find that the AGN fractions can be described by}
\begin{equation}
\log \fduty \approx -0.59 + \log \sfr
\label{eq:f001_vs_sfr}
\end{equation}\
and
\begin{equation}
\log \fbright \approx -1.34 + \log \sfr
\label{eq:f01_vs_sfr}
\end{equation}
while the average specific accretion rate is given by
\begin{equation}
\log \avsar = -3.68 + \log \sfr
\label{eq:avsar_vs_sfr}
\end{equation}
which we indicate by the dashed, dot-dashed and long-dashed lines in Figure~\ref{fig:fagn_vs_sfr}. 
We note that Equations~\ref{eq:f001_vs_sfr} and \ref{eq:f01_vs_sfr} do not provide a statistically acceptable fit to the $z<2.5$ data ($\tilde \chi^2 \approx 11.35$ and 5.05 for the $\sar>0.01$ and $\sar>0.1$ limits, respectively), indicating that there is an intrinsic scatter in this relation or significant additional systematic dependencies on stellar mass, redshift or some other confounding variable.
\upd{Indeed, systematic shifts in \Fduty\ as a function of redshift away from the basic 1:1 relation (black dashed line) are clearly seen in  Figure~\ref{fig:fagn_vs_sfr}~(top). 
The trend with SFR is somewhat tighter for \Fbright\ and does not exhibit the same level of redshift-dependent offset from the overall trend.
}
\mupd{For the average specific accretion rate, we find $\chi^2=37.1$ for $\nu=23$ degrees of freedom, thus the correlation is statistically acceptable ($\tilde \chi^2=1.61$), partly due to the large uncertainties in the measurements of \Avsar. 
The increase of the SFR moving \emph{along} the main sequence or to higher redshift (as the main sequence evolves) generally results in a higher AGN fraction and a higher average specific accretion rate, explaining both the stellar mass and redshift dependence of these quantities in star-forming galaxies.
}

This overall trend is consistent with our suggestion in \PaperII\ that the incidence of AGN in star-forming galaxies may be related to the availability of cold gas, which is also expected to determine the SFR and the evolution of the main sequence.
We also note that it is difficult to determine whether the correlation \emph{at a given redshift} is driven by a dependence on stellar mass or SFR , given our binning scheme and that these two quantities are correlated within the star-forming galaxy population. 
However, approximately the \emph{same} correlations with SFR appear to hold over a wide range in redshift, where both the average SFRs and incidence of AGN at a fixed \Mstel\ increase by a factor $\sim10$. Thus, the absolute value of the SFR may be the key parameter that determines the AGN content of star-forming galaxies. 
We explore these issues further in Section~\ref{sec:discuss_massvssfr} below. 

At higher redshifts ($z>2.5$), our measurements of \Fduty\ and \Fbright\ tend to lie below the relations given by Equations~\ref{eq:f001_vs_sfr} and \ref{eq:f01_vs_sfr} (shown by the dashed and dot-dashed lines in Figure~\ref{fig:fagn_vs_sfr}).
The AGN fraction cannot exceed 100~per~cent, thus these correlations cannot extend to arbitrarily high SFRs. 
At such high redshifts the accretion rate distributions, \Psar, appear to narrow close to the approximate Eddington limit ($\sar=1$).
\mupd{Thus, the average specific accretion rates continue to increase toward high redshift (as seen in the bottom panel of Figure~\ref{fig:fagn_vs_sfr}), even if this is not associated with a corresponding increase in the AGN fraction.} 
At such high redshifts, the incidence of AGN may be limited due to the self-regulation of black hole growth, when copious gas is available (see figure~9 of \PaperII\ and discussion therein).

\begin{figure}
\includegraphics[width=0.95\columnwidth,trim=10 50 40 30]{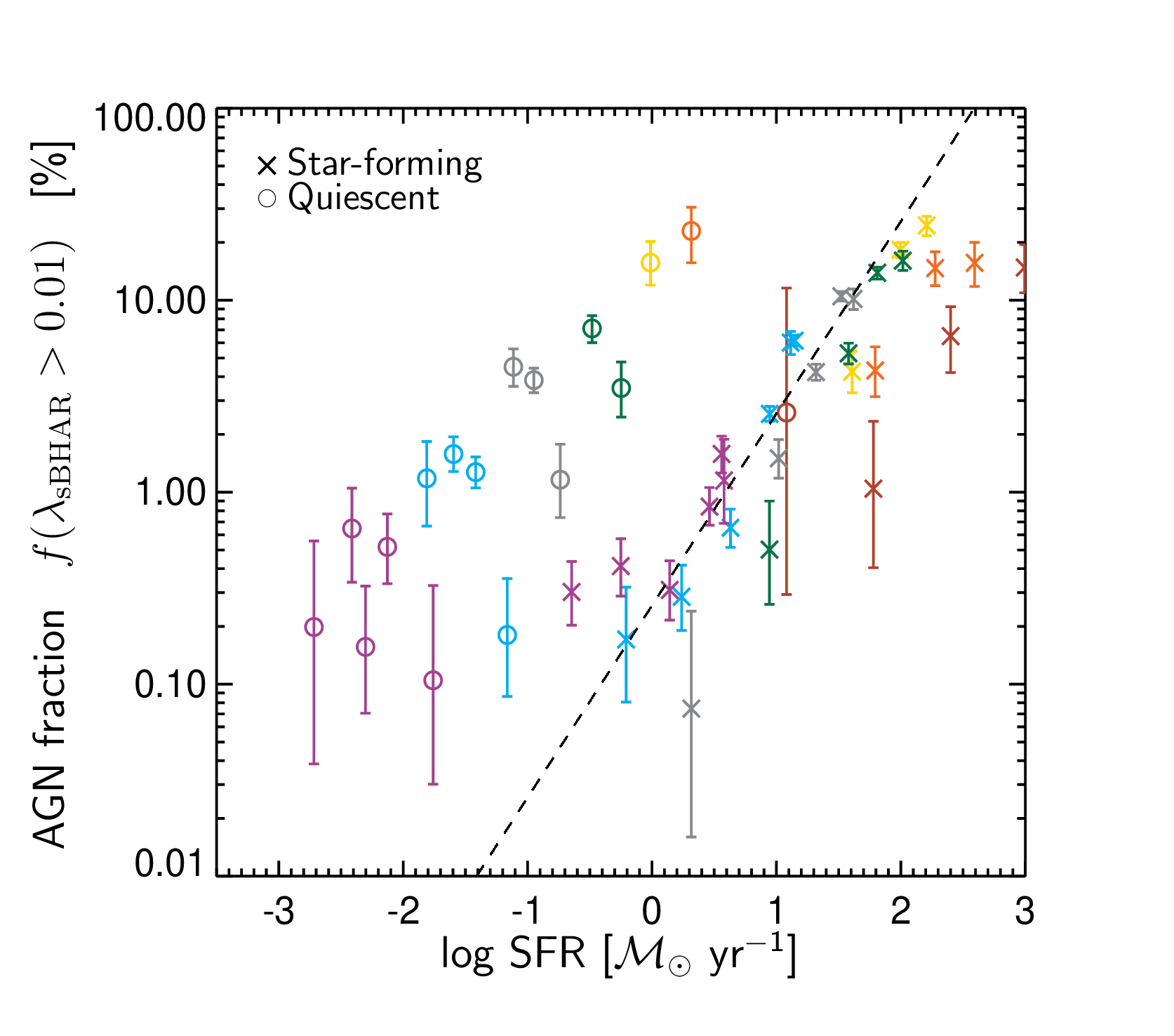}
\includegraphics[width=0.95\columnwidth,trim=10 50 40 10]{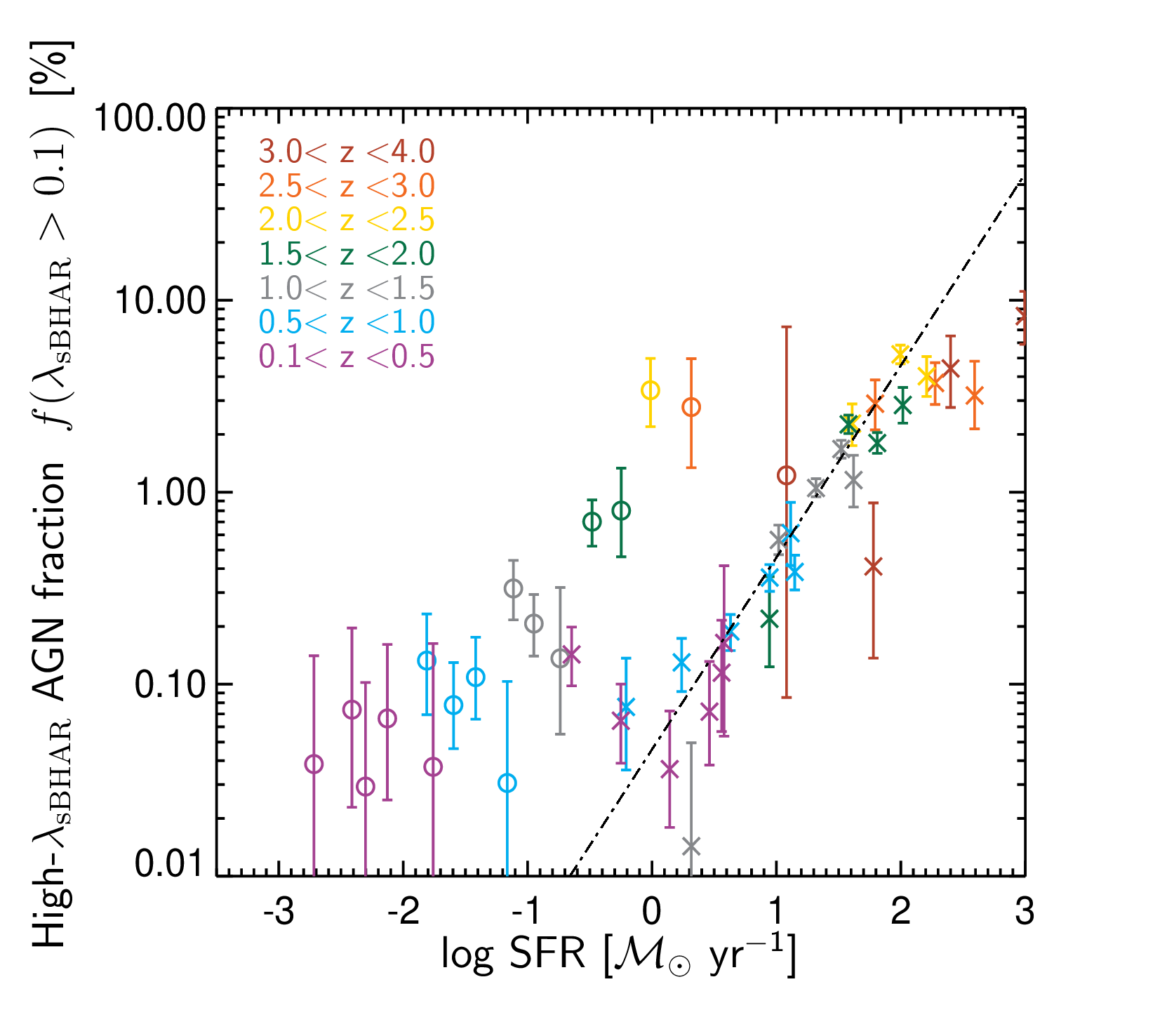}
\includegraphics[width=0.95\columnwidth,trim=10 30 40 10]{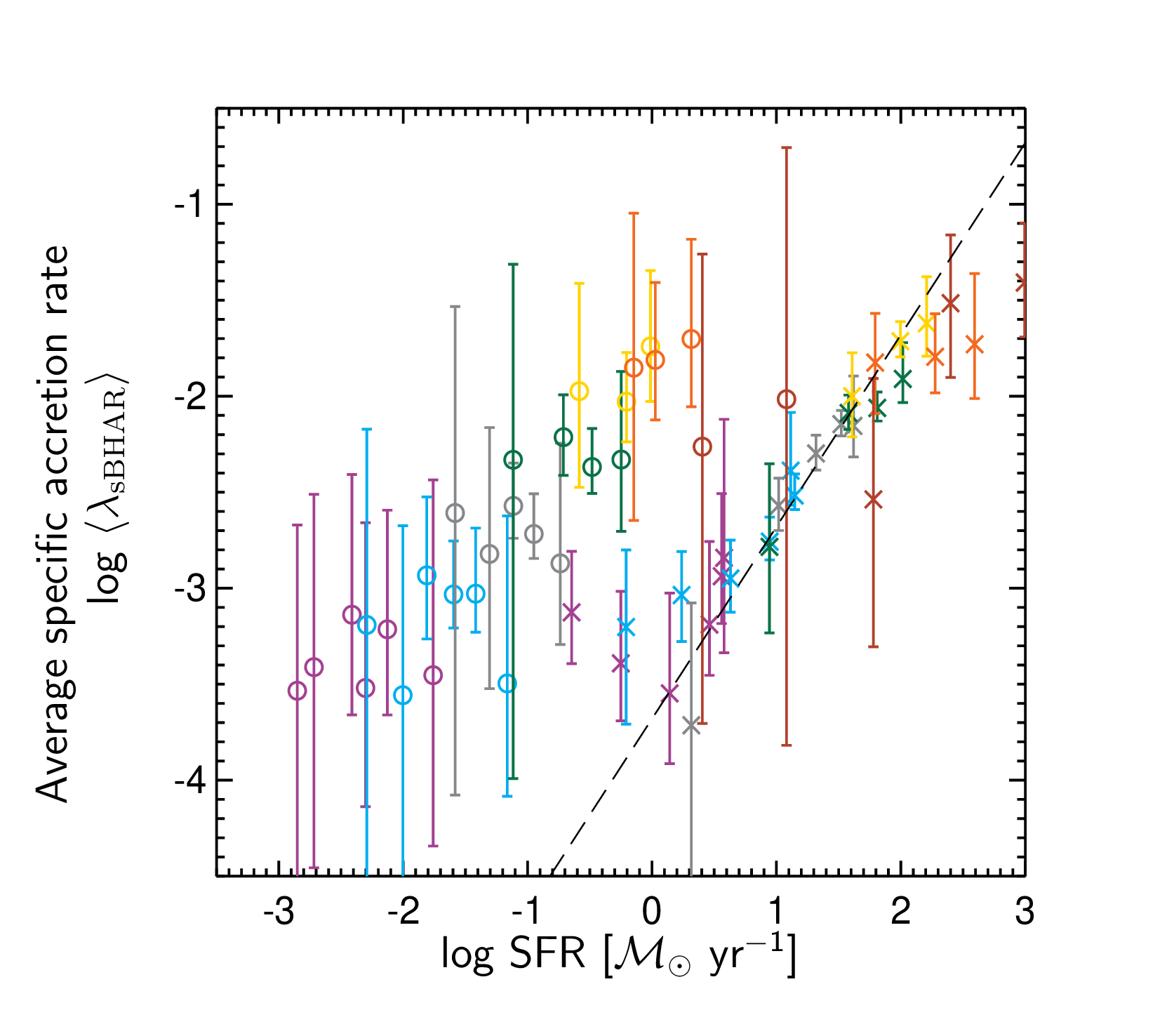}
\caption{
\mupd{
AGN fraction (\emph{top panel}), high-\Sar\ AGN fraction (\emph{middle panel}) and average specific accretion rate (\emph{bottom panel}) as a function of SFR, comparing star-forming galaxies (crosses)} and quiescent galaxies (circles). 
\upd{Each data point for a given galaxy type and redshift (as indicated by the symbol type and colour) corresponds to a different stellar mass.}
The dashed, dot-dashed and long-dashed lines indicate the best-fitting 1:1 relationship with SFR for star-forming galaxies at $z<2.5$, as given by Equations~\ref{eq:f001_vs_sfr}, \ref{eq:f01_vs_sfr} and \ref{eq:avsar_vs_sfr}.
\upd{
Our measurements for quiescent galaxies lie substantially above these correlations, indicating an additional mechanism may fuel AGN in such galaxies.}
}
\label{fig:fduty_vs_sfr_onepanel_sfqu}
\end{figure}

\begin{figure}
\includegraphics[width=0.95\columnwidth,trim=10 50 40 30]{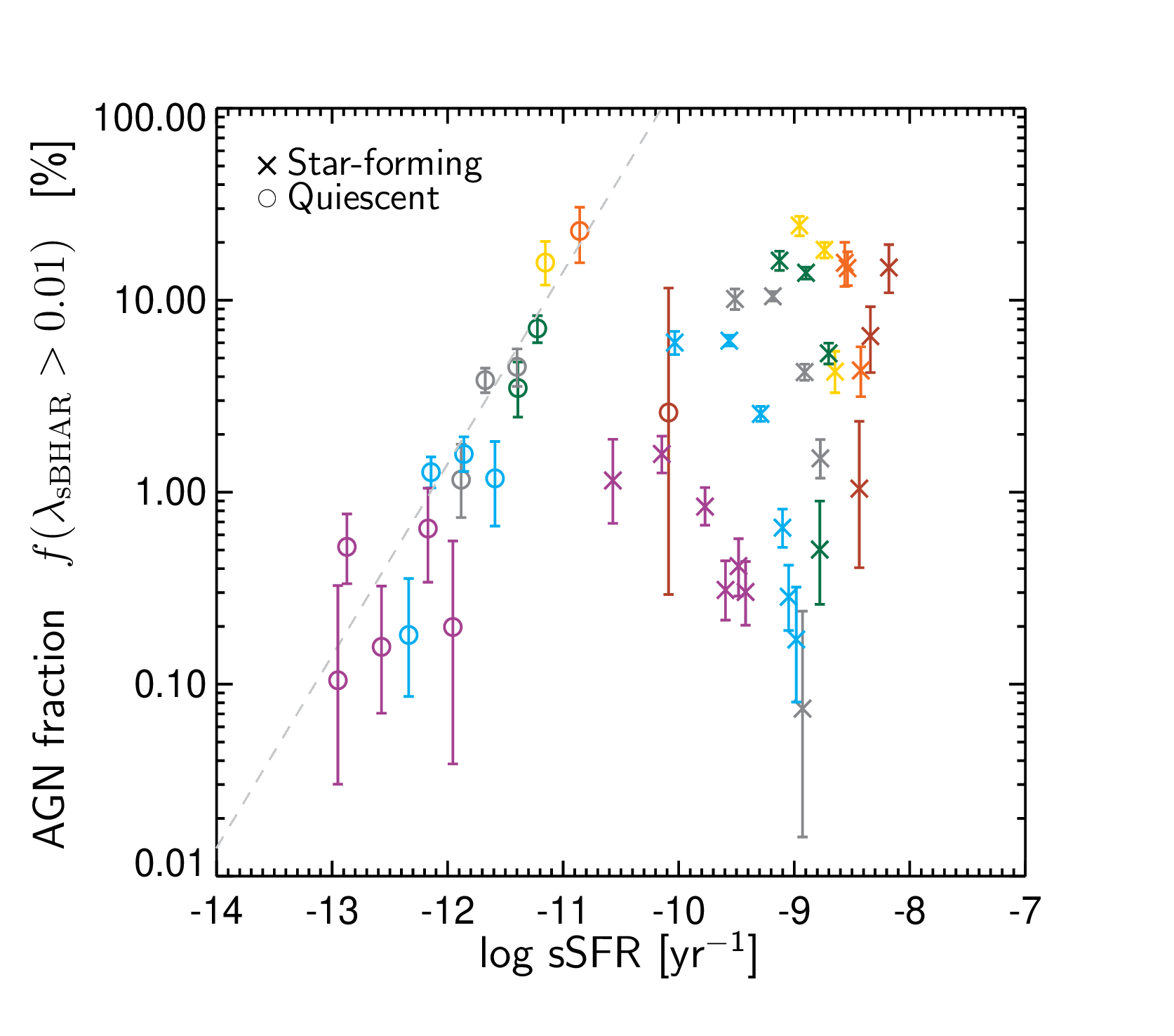}
\includegraphics[width=0.95\columnwidth,trim=10 50 40 10]{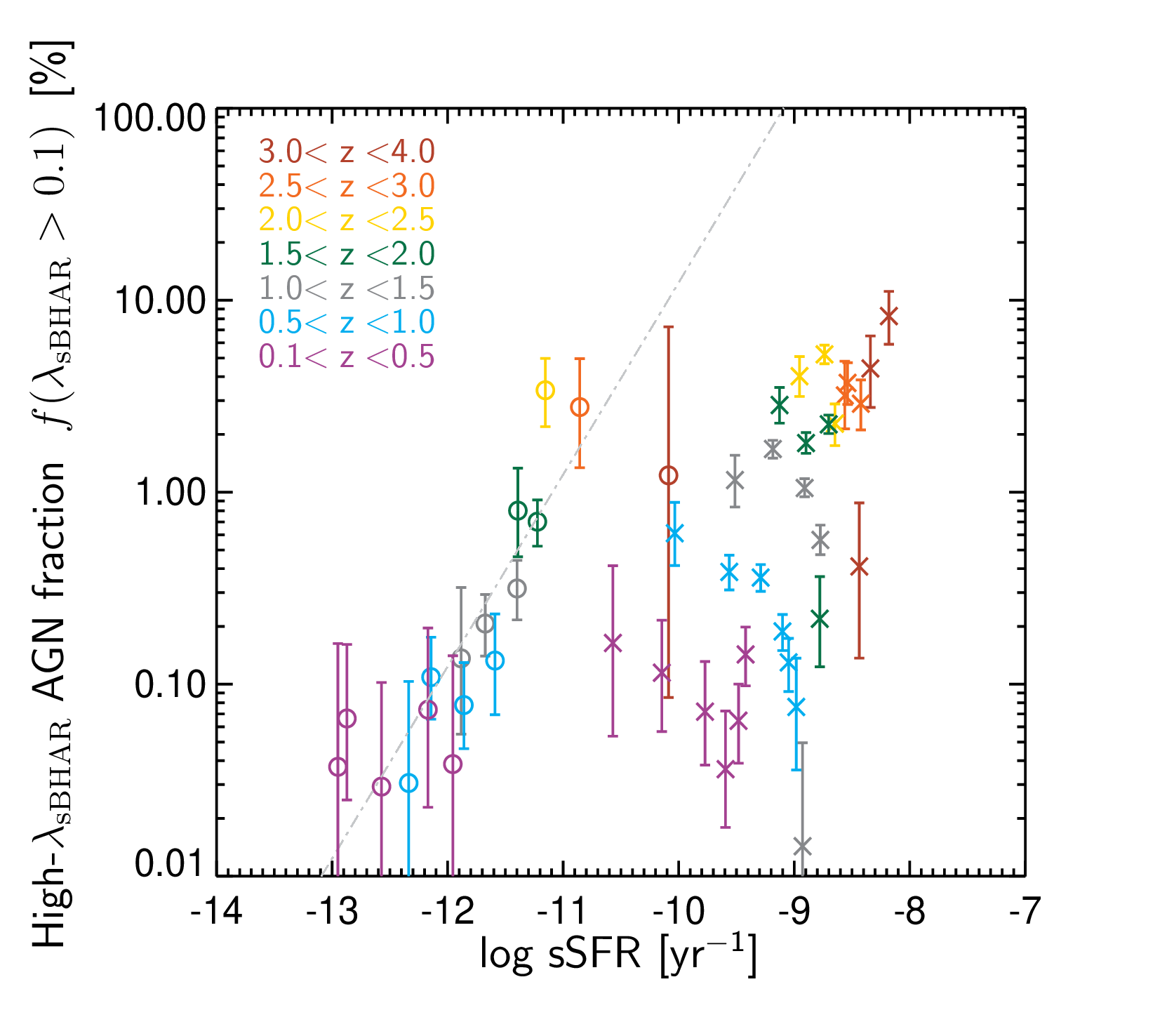}
\includegraphics[width=0.95\columnwidth,trim=10 30 40 10]{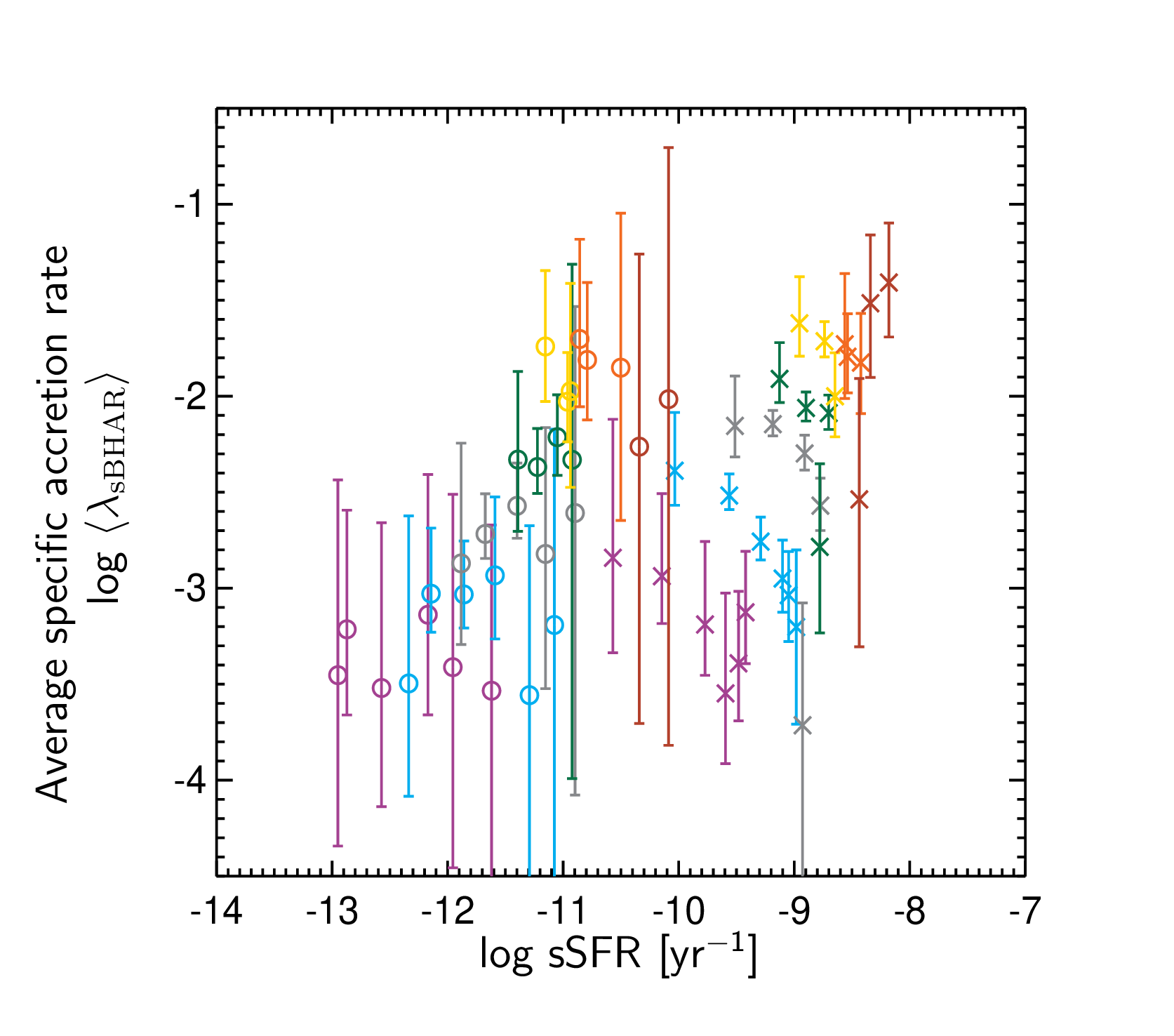}
\caption{
AGN fraction (\emph{top}), high-\Sar\ AGN fraction (\emph{middle}), and average specific accretion rate (\emph{bottom}) for star-forming and quiescent galaxies 
as presented in Figure~\ref{fig:fduty_vs_sfr_onepanel_sfqu} but 
shown as a function of the specific SFR (sSFR). 
The simple linear relation seen in the star-forming galaxies across all stellar masses and redshifts is no longer seen but there appears to be a tighter correlation between the AGN fraction or the high-\Sar\ AGN fraction with the sSFR for the quiescent galaxy samples (\upd{grey lines indicate a fit with a 1:1 slope}).
}
\label{fig:fduty_vs_ssfr_onepanel_sfqu}
\end{figure}

\upd{
In Figure~\ref{fig:fduty_vs_sfr_onepanel_sfqu} we compare our measurements of the incidence of AGN in star-forming galaxies (crosses) to the AGN incidence in quiescent galaxies (open circles) over the same range of stellar mass and redshift, where we have expanded the dynamic range of the $x$-axis compared to Figure~\ref{fig:fagn_vs_sfr} to accommodate the wider range of SFRs.}
We find that the AGN fractions in quiescent galaxies are comparable to or exceed the AGN fractions within star-forming galaxies (crosses) at a given redshift, despite the significantly lower SFRs of the quiescent galaxies.
The measured AGN fractions in quiescent galaxies all lie above the trends described by Equations~\ref{eq:f001_vs_sfr} and \ref{eq:f01_vs_sfr}. 
\mupd{Our measurements of the average specific accretion rates in quiescent galaxies are subject to large uncertainties but generally also lie above the trend identified in star-forming galaxies and given by Equation~\ref{eq:avsar_vs_sfr}.}
This result holds when comparing lower mass star-forming galaxies ($\log \mstel/\msun <10$), which have lower average SFRs, with higher mass quiescent galaxies at a fixed redshift.

Higher redshift quiescent galaxies that have \emph{equivalent} SFRs to lower redshift star-forming galaxies also exhibit an enhanced AGN fraction.
This result is consistent with our conclusions in \PaperII, where we suggested that a different physical mechanism (such as stellar mass loss) may be responsible for triggering and fuelling AGN activity within quiescent galaxies, in contrast to star-forming galaxies where the availability of cold gas may drive both star formation and the stochastic fuelling of AGN \citep[see also][]{Georgakakis11b,Georgakakis14}.
\upd{Thus, the SFR is \emph{not} the only property of a galaxy that determines the incidence of AGN, especially when considering galaxies that are \emph{not} on the main sequence (we explore this further in Section~\ref{sec:fagn_vs_mainseq} below).}

In Figure~\ref{fig:fduty_vs_ssfr_onepanel_sfqu} we also show the dependence of \Fduty, \Fbright, and \Avsar\ on the \emph{specific} SFR (sSFR, i.e.~SFR$/\mstel$) in both star-forming and quiescent galaxies. 
Adopting the sSFR breaks the simple one-to-one correlation seen in the star-forming galaxies.
\rone{At a fixed redshift the AGN fractions \emph{decrease} with increasing sSFR and the measurements at different redshifts do not lie on the same relation, in contrast to Figure~\ref{fig:fduty_vs_sfr_onepanel_sfqu}. 
Thus, it appears that the absolute value of the SFR is key in determining the AGN content of star-forming galaxies, rather than the relative amount of star-formation compared to the stellar mass of the galaxy.
For quiescent galaxies this pattern is reversed. 
In Figure~\ref{fig:fduty_vs_sfr_onepanel_sfqu} there are indications that the AGN fraction \emph{decreases} with increasing SFR at a fixed redshift. 
Across the different redshift bins there is a very broad correlation, mostly driven by the higher SFRs of quiescent at higher redshifts.
However, in Figure~\ref{fig:fduty_vs_ssfr_onepanel_sfqu} there are tentative indications of a positive correlation between the AGN fractions and sSFRs of quiescent galaxies \emph{at a fixed redshift}, at least at $z=0.5-1.0$ (light blue circles), $z=1.0-1.5$ (grey circles) and $z=1.5-2.0$ (green circles); we have only a single measurement in each of the higher redshift bins and the pattern is less clear in the $z=0.1-0.5$ redshift bin.
The same roughly 1:1 correlations between sSFR and the AGN fractions appear to hold across the different redshifts (as indicated by the light grey dashed and dot-dashed lines in Figure~\ref{fig:fduty_vs_ssfr_onepanel_sfqu}), 
although we note that quiescent galaxies at the same redshift tend to have a relatively narrow range of sSFR and thus the observed correlations between AGN fractions and sSFR may be partly driven by redshift effects.
Nonetheless, this trend between sSFR and AGN fraction is consistent with our suggestion that a different physical mechanism (e.g. stellar mass loss) fuels AGN activity in quiescent galaxies, as opposed to the stochastic accretion of cold gas that we suggest drives the correlation between the absolute SFR and the incidence of AGN in star-forming galaxies.}
\mupd{No clear trend is seen between the average specific accretion rate and sSFR in quiescent galaxies (bottom panel of Figure~\ref{fig:fduty_vs_ssfr_onepanel_sfqu}).}

\section{The incidence of AGN as a function of star formation rate relative to the main sequence of star formation}
\label{sec:fagn_vs_mainseq}

\begin{figure}
\includegraphics[width=\columnwidth,trim=40 25 50 10]{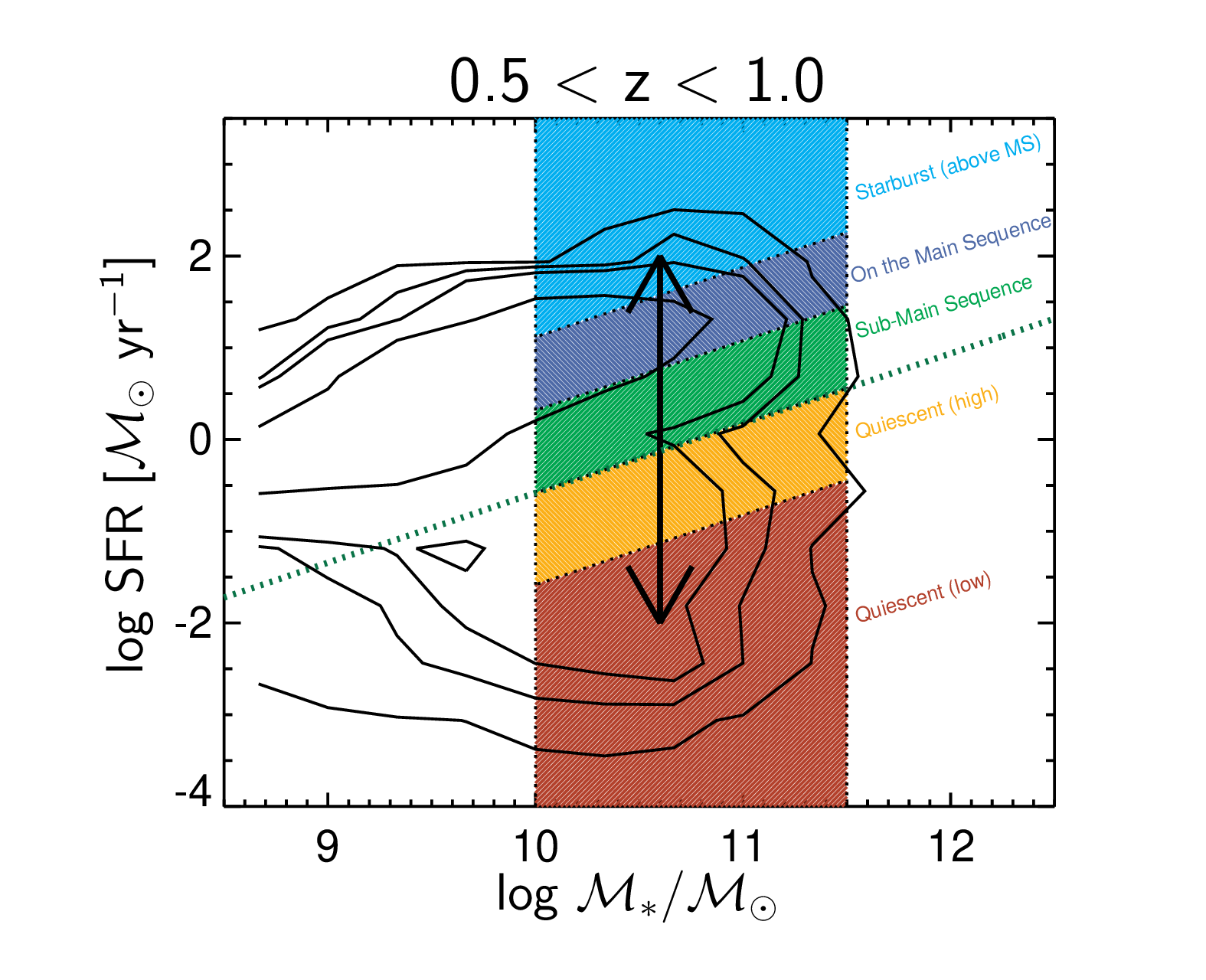}
\caption{
The distribution of SFRs as a function of \Mstel\ for our galaxy sample in a single redshift bin, as in Figure~\ref{fig:sfr_vs_mstel_bins1}, but now indicating our single stellar mass bin ($10.0<\log \mstel/\msun<11.5$) and five classifications based on their SFRs relative to the star-forming main sequence, \sfratio, used to divide the galaxy population and study the incidence of AGN in Section~\ref{sec:fagn_vs_mainseq}.
}
\label{fig:sfr_vs_mstel_bins2}
\includegraphics[width=\columnwidth,trim=40 20 60 0]{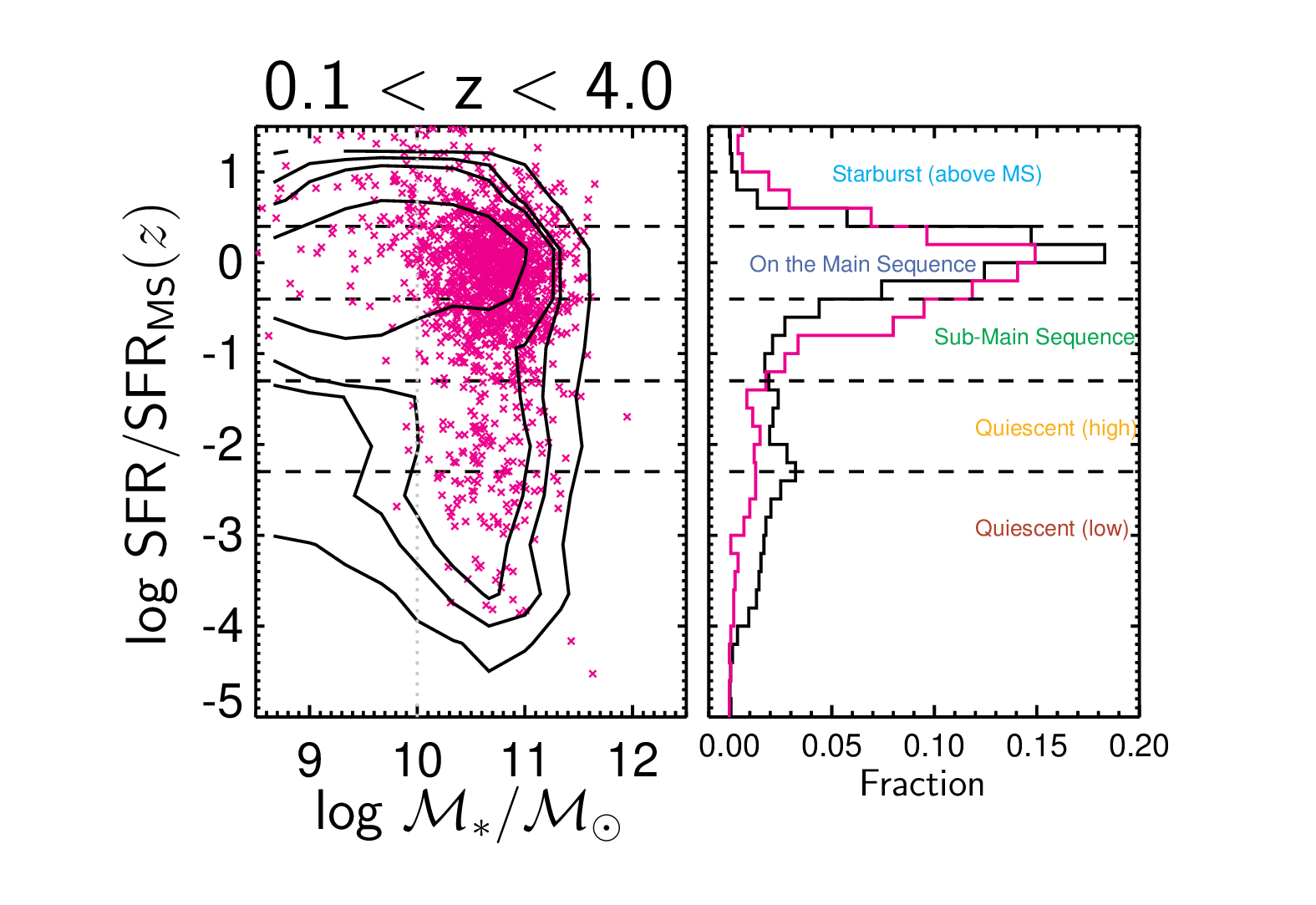}
\caption{Measurements of \sfratio, the star formation rates relative to the redshift-dependent main sequence of star formation for galaxies over our full redshift range, $0.1<z<4$. 
Black contours enclose 68, 90, 95 and 99 per cent of galaxies.
Pink crosses indicate hard X-ray detections (i.e. AGN). 
The right panel shows the overall distributions of \sfratio\ for the galaxies (black) and X-ray detected AGN (pink) with stellar masses $\mstel>10^{10}\msun$.
We divide the galaxy population into five classifications, as indicated (see also Figure~\ref{fig:sfr_vs_mstel_bins2}).
Compared to the galaxy population, X-ray detected AGN have a broader distribution of \sfratio\ and thus exhibit a slight preference to be found in galaxies that lie below the main sequence. 
Our measurements of AGN fractions (see Figures~\ref{fig:fduty_vs_sfr_onepanel_msbins} and \ref{fig:fduty_vs_z_fivebins}) investigate these trends in a more robust manner.
}
\label{fig:ensfr_vs_mass}
\end{figure}

In this section, we explore the relationship between the SFR of galaxies and their AGN content in greater detail, dividing our galaxy samples according to their SFRs relative to the main sequence of star formation, \sfratio\ \citep[see also][]{Azadi15}.
Figure~\ref{fig:sfr_vs_mstel_bins2} illustrates how our \sfratio\ binning divides the galaxy sample in the \hbox{\Mstel-SFR} plane at a given redshift. 
This sub-division allows us to identify trends \emph{within} (as opposed to between) the star-forming and quiescent galaxy populations and provides greater insight into the relationship between the SFRs of galaxies and the incidence of AGN activity within these galaxies.
We adopt the redshift-dependent main sequence measured in \PaperI\ and given by
\begin{equation}
\log \mathrm{SFR_\mathrm{MS}}(z) \mathrm{[\msun yr^{-1}]} = -7.6 + 0.76\log \dfrac{\mstel}{\msun} + 2.95 \log (1+z).
\label{eq:sfms}
\end{equation}
and then divide our galaxies into five samples:
\begin{enumerate}
\item
Star-forming galaxies with elevated SFRs compared to the main sequence, i.e. $\log  \mathrm{SFR}/\mathrm{SFR_{MS}}(z)>0.4$, hereafter referred to as ``Starburst" galaxies.
\item 
Those with SFRs that place them on the main sequence i.e. with \sfratio\ within $\pm0.4$~dex of the main sequence and corresponding to $\sim70$~per cent of the overall star-forming galaxy population, hereafter referred to as ``Main Sequence" galaxies.
\item 
Star-forming galaxies with SFRs that are lower than the bulk of the Main Sequence galaxies but above the level of a quiescent galaxy, i.e. with $-1.3<\log  \mathrm{SFR}/\mathrm{SFR_{MS}}(z)< -0.4$, hereafter referred to as ``Sub-MS" galaxies.
\item
Quiescent galaxies with higher SFRs, within the top $\sim$50~per~cent of the quiescent galaxy population, 
i.e. $-2.3<\log  \mathrm{SFR}/\mathrm{SFR_{MS}}(z)< -1.3$, hereafter referred to as ``Quiescent (high)" galaxies.
\item
Quiescent galaxies with lower SFRs, i.e. $\log  \mathrm{SFR}/\mathrm{SFR_{MS}}(z)< -2.3$, hereafter referred to as ``Quiescent (low)" galaxies.
\end{enumerate}

We adopt a single, relatively broad stellar mass bin ($10.0<\log \mstel/\msun<11.5$) to ensure we have sufficiently large samples of galaxies and can obtain reliable estimates of \Psar\ for each of the five galaxy populations.
We note that adopting a single, broad stellar mass bin will dilute any stellar mass dependence in \Psar\ or our derived quantities (AGN fractions and average specific accretion rates).
However, the median stellar mass changes by $<0.3$~dex between the five galaxy samples and our results will be dominated by galaxies at $\mstel \sim 10^{10-11}\msun$, where the majority of our galaxy sample lies and thus any underlying stellar-mass dependence is not expected to have a significant impact on our results.

\begin{figure*}
\includegraphics[width=\textwidth,trim=10 35 10 35]{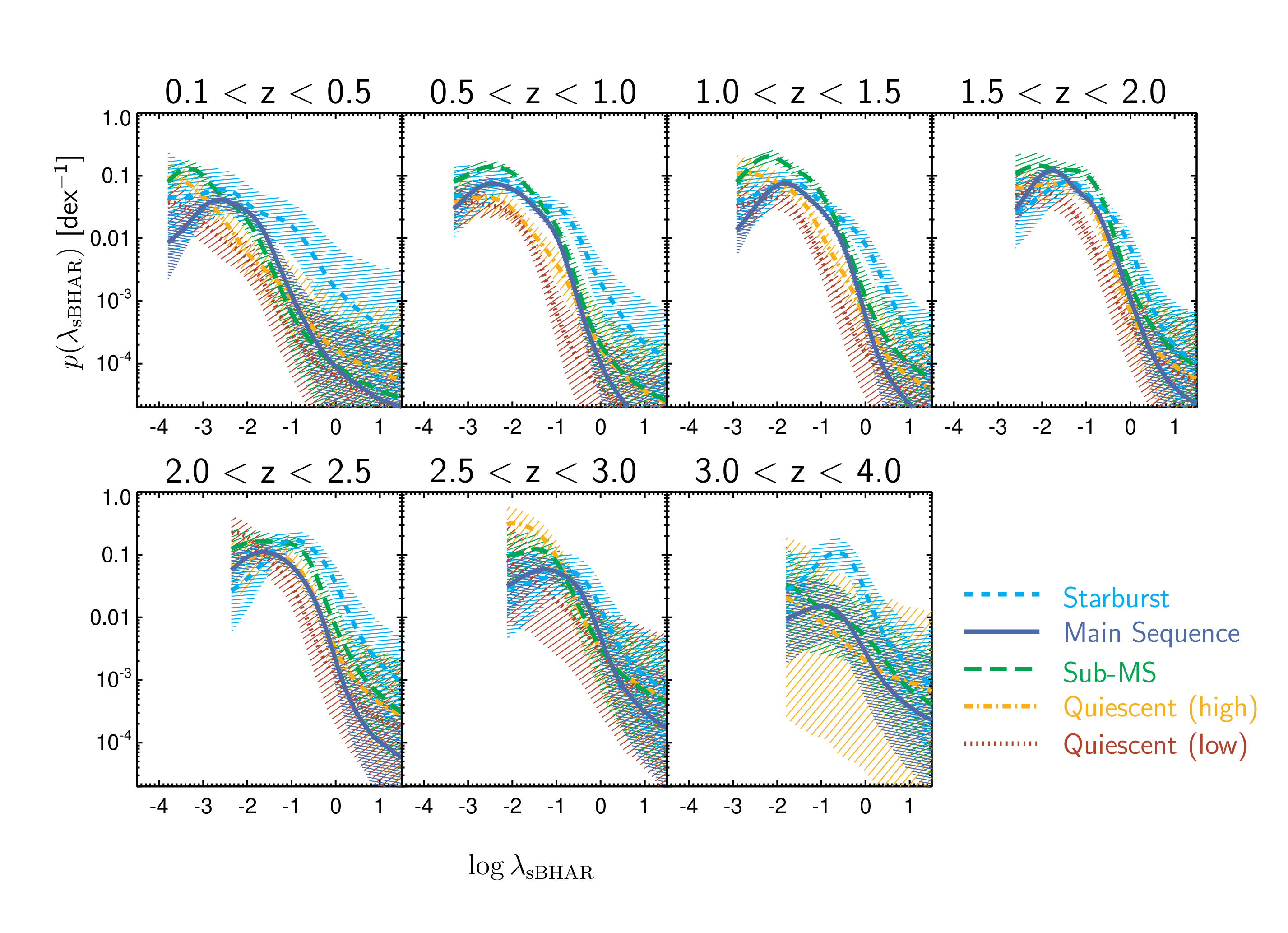}
\caption{
Probability distributions of specific black hole accretion rates, \Psar, within our five galaxy populations based on \sfratio\ (see Figure~\ref{fig:sfr_vs_mstel_bins2}).
We consider galaxies with stellar masses in the range $10.0<\log \mstel/\msun<11.5$ and with redshifts within the indicated limits. 
The coloured lines represent our best estimates of \Psar\ for each class, as indicated, and the shaded regions indicate the 90 per cent confidence intervals, as determined by our flexible Bayesian modelling. 
}
\label{fig:pledd_fivebins}
\end{figure*}

By measuring SFRs of galaxies normalized relative to the main sequence (at the redshift of the galaxy) we are able account for the broad relation between stellar mass and SFR, whereby a typical star-forming galaxy with a higher \Mstel\ also has a correspondingly higher SFR (which is a significant effect across our broad stellar mass bin). 
We also account for the overall redshift evolution of the galaxy population, whereby the typical SFRs of star-forming galaxies are higher at higher redshifts. 
With this approach, all our SFR measurements are relative to the bulk of the galaxy sample and thus we will not be significantly affected by potential systematic errors in our SFR estimates. 
Furthermore, tracking the global evolution of the galaxy population in such a manner ensures that we retain a sufficient sample of galaxies in each of our \sfratio\ bins over our full redshift range. 
At a fixed redshift, an increase in \sfratio\ directly tracks an increase in the SFR.
However, the \emph{absolute} SFRs of a given galaxy sample will vary between the different redshift bins.

Figure~\ref{fig:ensfr_vs_mass} shows the distribution of \sfratio\ for our galaxy sample.
The right panel compares the distributions of \sfratio\ for our galaxy sample (black) and the X-ray detected sample (pink), both subjected to our stellar mass limits of $10.0 < \log \mstel/\msun <11.5$. 
The distributions indicate a slight preference for X-ray AGN to be identified within galaxies with SFRs that are below the main sequence \citep[see also e.g.][]{Aird12,Shimizu15,Mullaney15b}.
 \upd{We also note a slight enhancement in the number of AGN above the main sequence and a more rapid fall-off in the number of AGN in the Quiescent~(high) and Quiescent~(low) regimes compared to the full galaxy sample.}
In the remainder of this section we explore these trends in a more robust manner (accounting for selection effects) 
by measuring the distributions of accretion rates, robust AGN fractions and average specific accretion rates within the different galaxy populations.

In Figure~\ref{fig:pledd_fivebins} we present measurements of \Psar\ for the five galaxy populations at different redshifts.	 
We are able to identify a number of trends, in common at all redshifts out to at least $z=2$. 
First, we note the relatively broad distributions of \Sar\ within each galaxy population and at all redshifts, likely reflecting the stochastic nature of AGN activity within a given galaxy sample.
In general, \Psar\ decreases toward higher values of \Sar, indicating that higher accretion rate AGN are rarer within a given galaxy sample than lower accretion rate AGN. 
\upd{For Main Sequence galaxies, we also see clear evidence of an additional turnover in \Psar\ towards lower 
accretion rates, 
indicating a preferential rate of accretion in such galaxies and a lack of lower accretion rate sources.
In Starburst galaxies, \Psar\ appears shifted toward higher \Sar, generally indicating a higher probability of such galaxies hosting an AGN, although the turnover at lower \Sar\ may be stronger than in Main Sequence galaxies (especially at $z\gtrsim2$).
Sub-MS galaxies do not exhibit an overall shift toward higher accretion rates, but \Psar\ is generally higher than in Main Sequence galaxies, especially at moderate accretion rates ($-3\lesssim \log \sar \lesssim -1$).
Moving to lower SFRs, into the Quiescent~(high) sample, we find that the amplitude of \Psar\ drops again, becoming comparable to or below \Psar\ for Main Sequence galaxies and showing no evidence of a low-\Sar\ turnover. 
}
The quiescent galaxy samples appear to contain progressively weaker, lower accretion rate AGN as the SFRs reduce i.e. moving from the star-forming galaxy samples to the Quiescent (high) and then Quiescent (low) populations. 
Nevertheless, the incidence of the lowest accretion rate AGN ($\sar \lesssim 0.01$) within the quiescent galaxy samples may exceed the incidence of such sources in the higher SFR populations.

\begin{figure}
\includegraphics[width=0.95\columnwidth,trim=10 50 40 30]{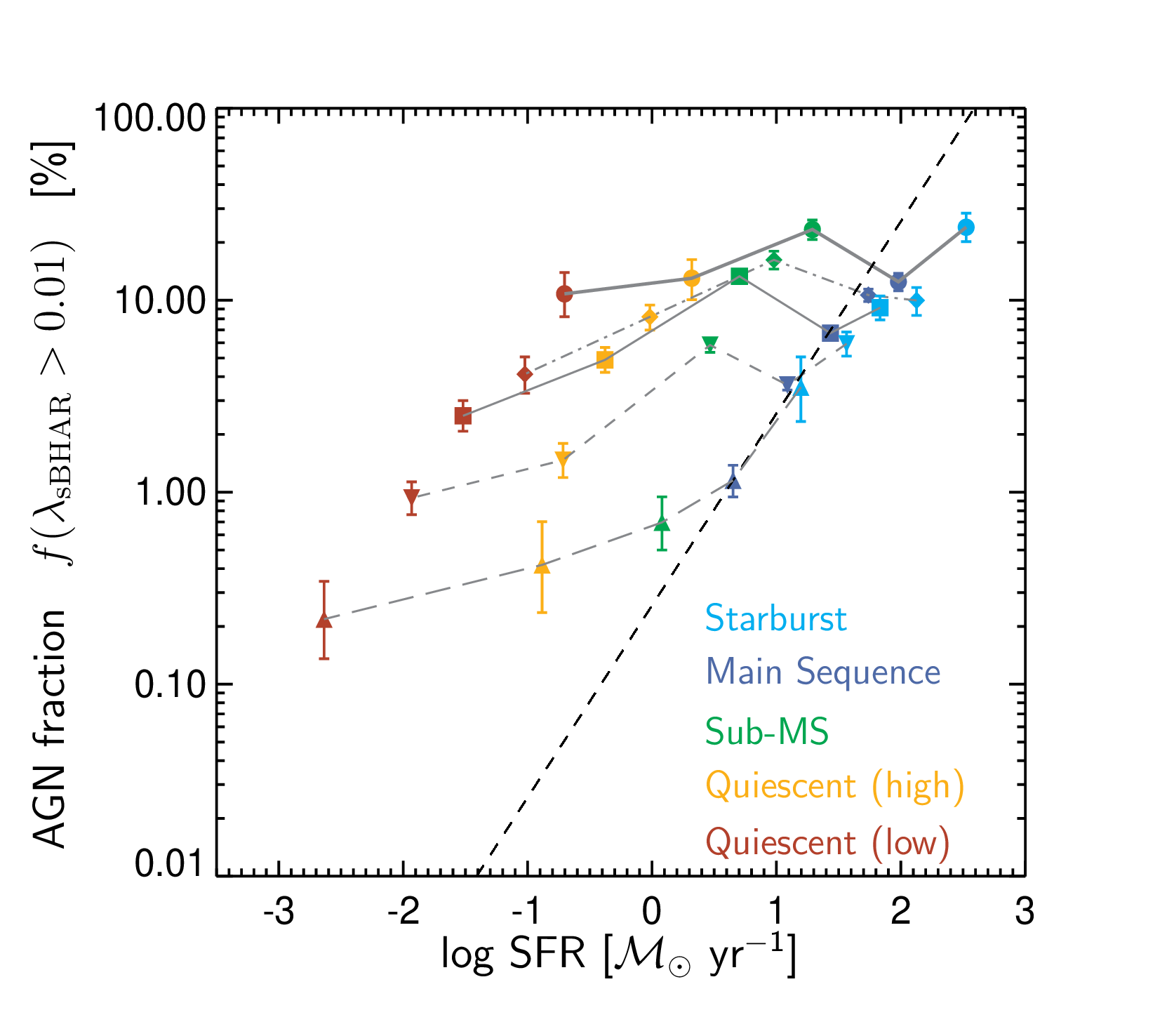}
\includegraphics[width=0.95\columnwidth,trim=10 50 40 10]{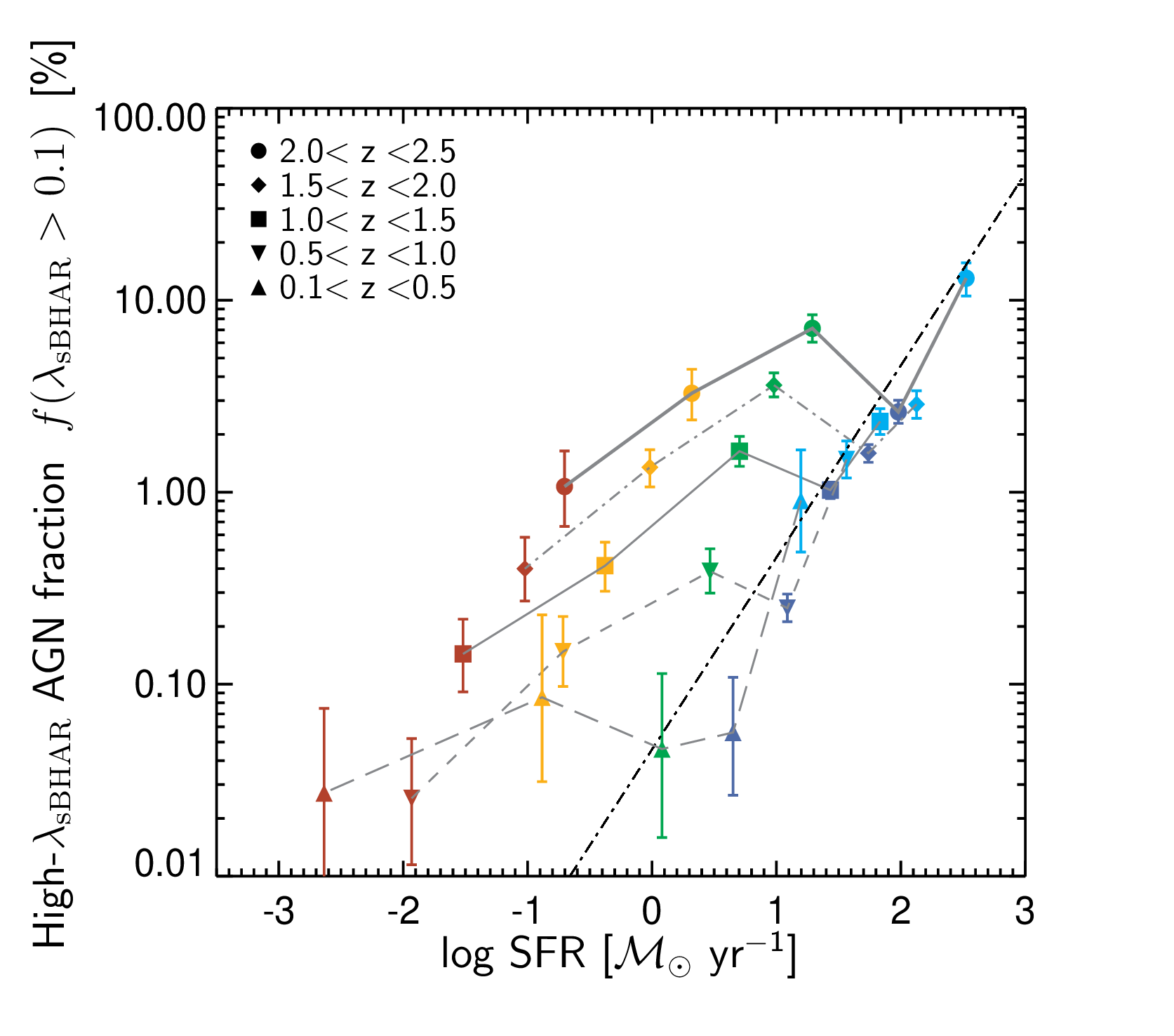}
\includegraphics[width=0.95\columnwidth,trim=10 30 40 10]{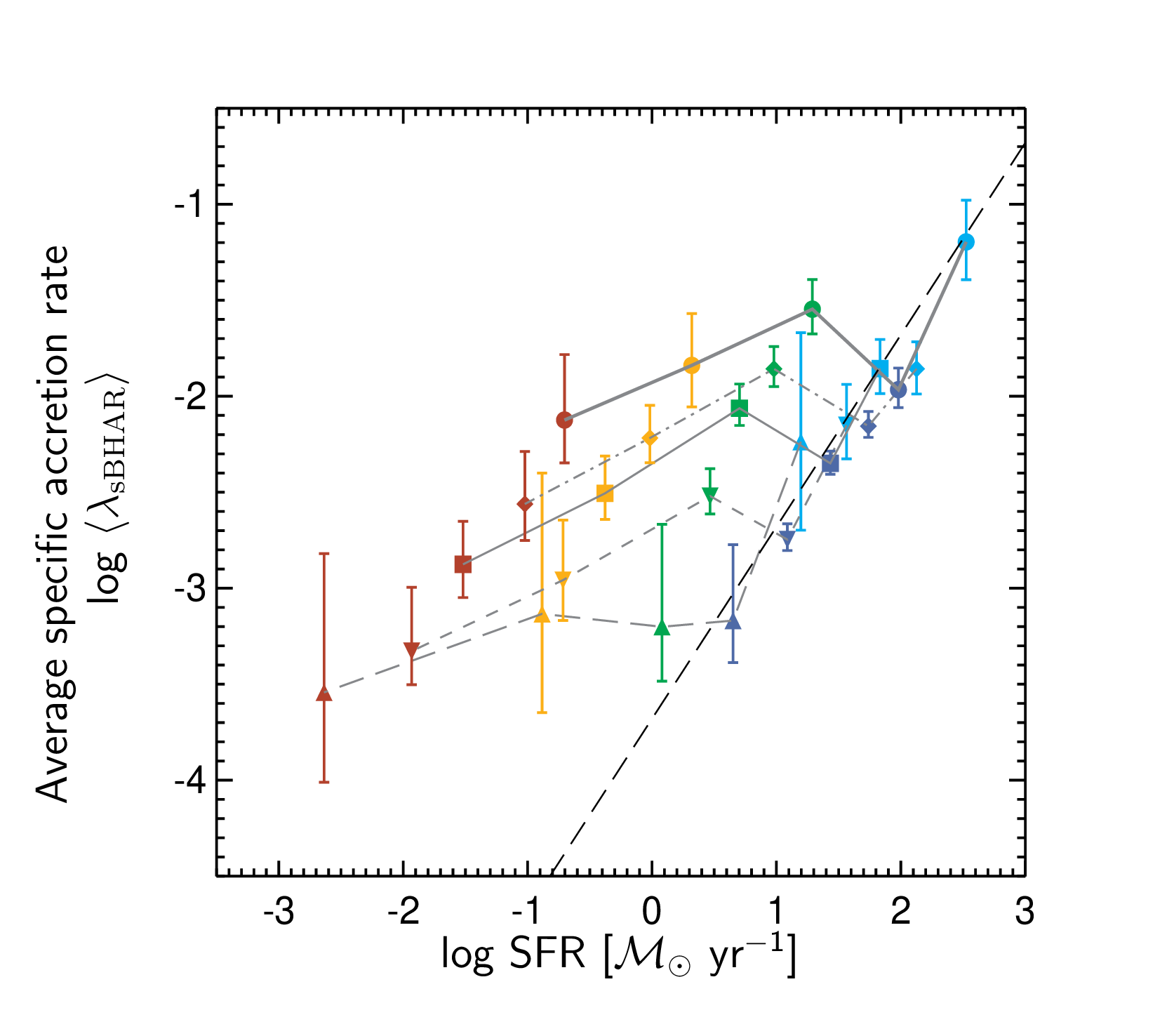}
\caption{
Measurements of the AGN fraction (\emph{top}), high-\Sar\ AGN fraction (\emph{middle}) and average specific accretion rate (\emph{bottom})  as a function of SFR for a single $10<\log \mstel/\msun<11.5$ mass bin and dividing the galaxy sample into five populations based on \sfratio, moving \emph{across} the main sequence of star formation at a given redshift.
Colours indicate the galaxy classification based on \sfratio\ while symbol types indicate the redshift.
Grey lines connect measurements at the same redshift. 
The dashed, dot-dashed and long-dashed black lines reproduce the linear relations with SFR found in Section~\ref{sec:fagn_sfandqu} for star-forming galaxies over a wide range of \Mstel\ and $z$ (see Equation~\ref{eq:f001_vs_sfr}, \ref{eq:f01_vs_sfr} and \ref{eq:avsar_vs_sfr} ).
}
\label{fig:fduty_vs_sfr_onepanel_msbins}
\end{figure}

\begin{figure}
\includegraphics[width=0.95\columnwidth,trim=10 50 40 30]{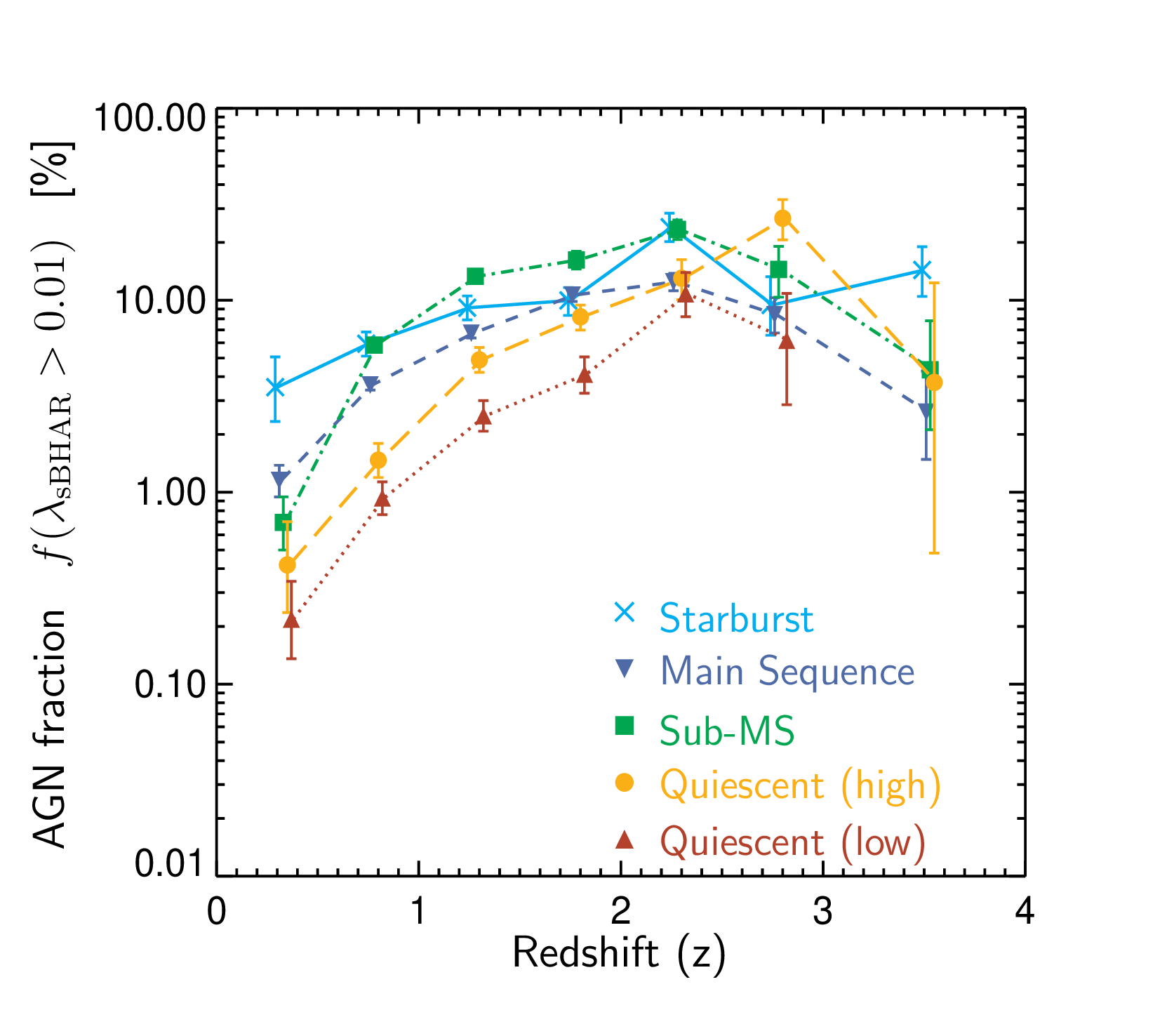}
\includegraphics[width=0.95\columnwidth,trim=10 50 40 10]{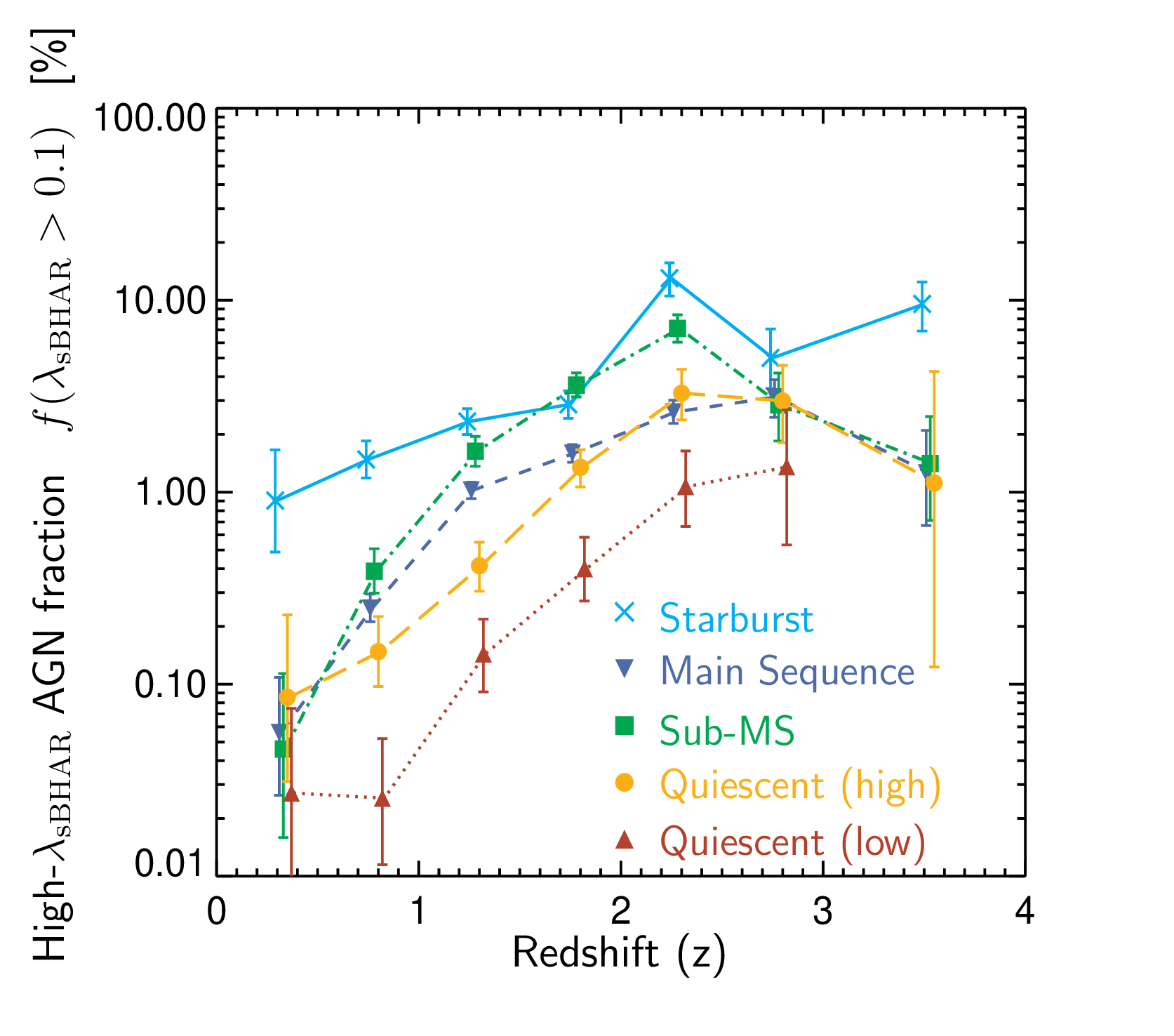}
\includegraphics[width=0.95\columnwidth,trim=10 30 40 10]{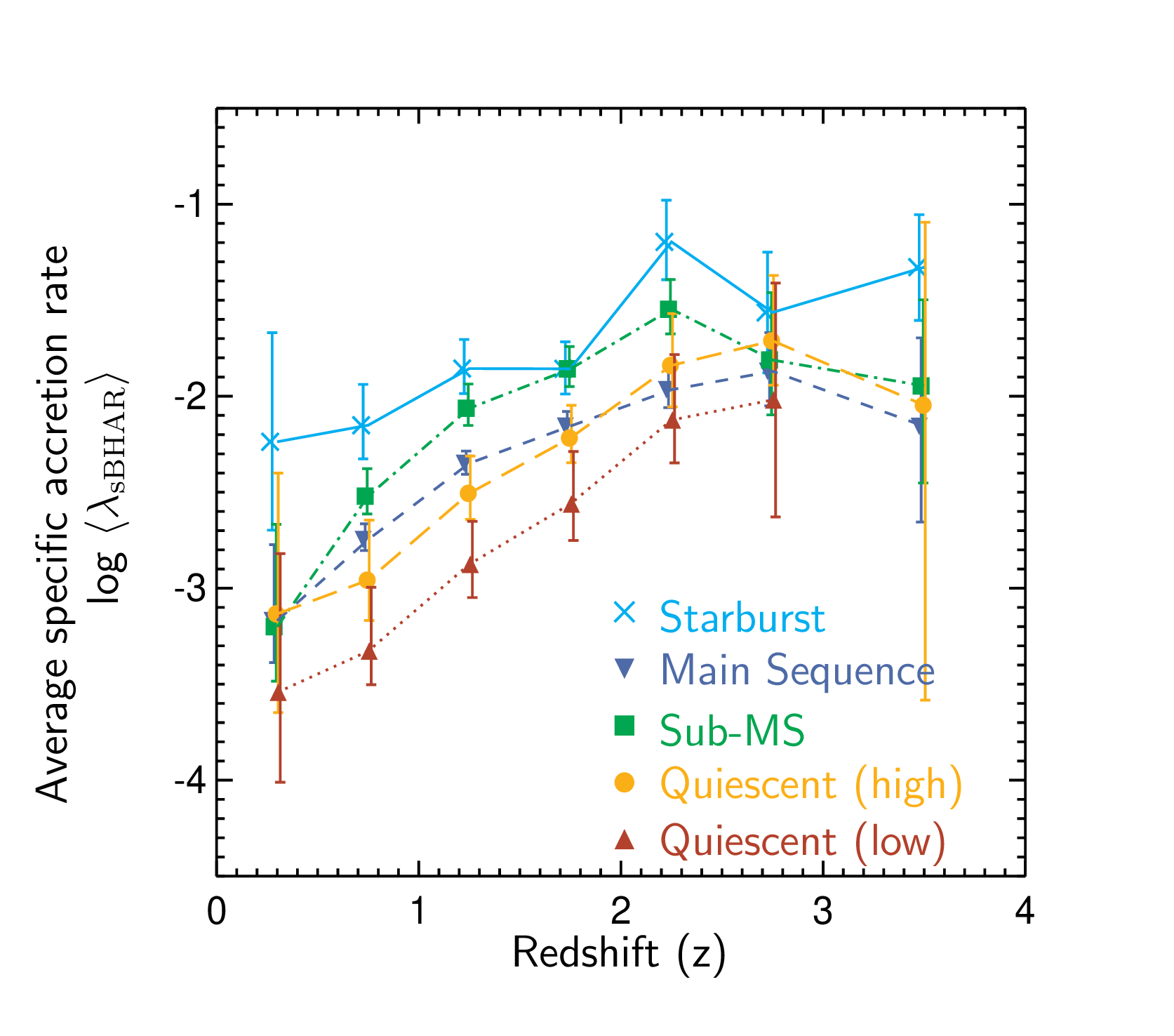}
\caption{
Measurements of the AGN fraction (\emph{top}), high-\Sar\ AGN fraction (\emph{middle}) and average specific accretion rate (\emph{bottom}) as a function of redshift for our five galaxy classifications based on \sfratio, as indicated by the symbol colours and types.
}
\label{fig:fduty_vs_z_fivebins}
\end{figure}

In Figure~\ref{fig:fduty_vs_sfr_onepanel_msbins}
we present measurements of the AGN fraction, \Fduty, the high-\Sar\ AGN fraction, \Fbright, and the average specific accretion rate, \Avsar, as a function of the SFR across our five galaxy populations (\upd{for clarity, we only present measurements at $z<2.5$}). \upd{These measurements allow us to summarize the trends seen in the measurements of \Psar\ and track how the incidence of AGN depends on the SFRs of galaxies at a fixed stellar mass and redshift, as we move \emph{across} the main sequence (cf.~Figure~\ref{fig:fduty_vs_sfr_onepanel_sfqu} where galaxies are binned in stellar mass and we thus probe changes in the AGN fraction \emph{along} the main sequence).}

\upd{
For Main Sequence galaxies, our measurements of \Fduty, \Fbright\ and \Avsar\ are in good agreement with the overall trends identified out to $z\approx2.5$ in Section~\ref{sec:fagn_sfandqu} (indicated by the dashed, dot-dashed and long-dashed black lines in Figure~\ref{fig:fduty_vs_sfr_onepanel_msbins}  and given by Equations~\ref{eq:f001_vs_sfr}, \ref{eq:f01_vs_sfr} and \ref{eq:avsar_vs_sfr}),
as expected given that these Main Sequence galaxies correspond to the bulk of the star-forming galaxy population. 
For galaxies on the Main Sequence, there is a strong correlation between the incidence of AGN and the SFR of the star-forming galaxies.
The global evolution of the galaxy population is also apparent, whereby the SFRs of Main Sequence galaxies are progressively higher at higher redshifts.}

\upd{
For Starburst galaxies, we find an enhancement in the incidence of AGN  compared to Main Sequence galaxies, seen most clearly in the high-\Sar\ AGN fractions and the average specific accretion rates (middle and bottom panels of Figure~\ref{fig:fduty_vs_sfr_onepanel_msbins}) and reflecting the excess of higher \Sar\ seen in our measurements of \Psar\ (Figure~\ref{fig:pledd_fivebins}). 
However, these measurements tend to lie close to the overall trends given by Equations~\ref{eq:f01_vs_sfr} and \ref{eq:avsar_vs_sfr} (dot-dashed and long-dashed black lines), indicating that the enhancement of high-\Sar\ AGN may be due to the higher SFRs of the Starburst galaxies.
}

\upd{
Comparing the Sub-MS and Main Sequence galaxy samples, we find a significant \emph{increase} (by a factor $\sim2-5$) in the incidence of AGN at a fixed redshift as SFR \emph{decreases} in Sub-MS galaxies compared to Main Sequence galaxies.
This enhancement is seen at $0.5<z<2.5$ for \Fduty, \Fbright, and \Avsar\ (top, middle and bottom panels of Figure~\ref{fig:fduty_vs_sfr_onepanel_msbins}, respectively).
The enhancement in Sub-MS galaxies is not seen in our lowest redshift range ($0.1<z<0.5$), although this may be due to the overall evolution of the distribution of accretion rates whereby \Psar\ shifts uniformly toward lower \Sar\ at lower $z$. 
An enhancement of the AGN content of Sub-MS galaxies, compared to Main Sequence galaxies, is seen in our lowest redshift range in our measurements of \Psar\ in Figure~\ref{fig:pledd_fivebins} but is only found at $\sar\lesssim 10^{-3}$ and is thus not reflected in the measurements shown in Figure~\ref{fig:fduty_vs_sfr_onepanel_msbins}.
}

\upd{
At lower SFRs, moving into the Quiescent (high) and Quiescent (low) populations at a given redshift, we see a mild decline in \Fduty, \Fbright\ and \Avsar\ that reflects the shift of \Psar\ toward lower \Sar. 
Nonetheless, the incidence of AGN in such samples remains significantly above the global relations between the AGN incidence and SFR for star-forming galaxies that was found in Section~\ref{sec:fagn_sfandqu} and is shown by the dashed, dot-dashed and long-dashed lines in Figure~\ref{fig:fduty_vs_sfr_onepanel_msbins}. 
Indeed, the measured relation between SFR and \Fduty, \Fbright\ and \Avsar\ shown in Figure \ref{fig:fduty_vs_sfr_onepanel_msbins},
at a given redshift is relatively flat, indicating that a reduction in SFRs within galaxies that lie below the main sequence is \emph{not} associated with a substantial reduction in the incidence of AGN.
These trends likely indicate different physical processes trigger and fuel AGN in galaxies on versus below the main sequence (see further discussion in Section~\ref{sec:discuss_belowMS}).
}

Finally, in Figure~\ref{fig:fduty_vs_z_fivebins} we show the evolution of \Fduty, \Fbright\ and \Avsar\ in our five galaxy populations as a function of redshift. 
The AGN fractions and average specific accretion rates in the Main Sequence, Sub-MS, Quiescent (high) and Quiescent (low) galaxy populations all undergo a broadly similar evolution, increasing rapidly by a factor $\gtrsim 10$ between $z\sim0.3$ and $z\sim2$. 
The enhanced incidence of AGN in Sub-MS galaxies compared to the Main Sequence population at a given redshift is clearly seen for $0.5<z<2$, as well as the decrease in incidence of AGN in the Quiescent (high) and Quiescent (low) populations. 
The evolution of \Fduty, \Fbright\ and \Avsar\ in Starburst galaxies, however, is more mild, increasing by a factor $\lesssim 3$ between $z\sim0.3$ and $z\sim2$, although we note 
the higher but potentially discrepant measurements at $z=2-2.5$. 
This different evolution means \Fduty\ in Starburst galaxies is higher than the other galaxy populations at $z=0.1-0.5$, comparable to Sub-MS galaxies at $z=0.5-1$, and lower than Sub-MS galaxies at $z=1-1.5$ and $z=1.5-2$. 
The high-\Sar\ AGN fraction in Starburst galaxies, however, is always higher or comparable to the other galaxy populations, reflecting the higher prevalence of the most rapidly accreting AGN within the Starburst population (see also Figure~\ref{fig:pledd_fivebins}). 
In Section~\ref{sec:discuss_starbursts} below we discuss whether these differences indicate that a more violent physical process, such as galaxy mergers, are responsible for increasing the incidence of high-\Sar\ AGN within Starburst galaxies.

\section{Discussion}
\label{sec:discuss}

In this section, we discuss the main results from Sections~\ref{sec:fagn_sfandqu} and \ref{sec:fagn_vs_mainseq} above. 
We first discuss the overall correlation between the incidence of AGN and the SFR in star-forming galaxies, found in Section~\ref{sec:fagn_sfandqu}, \upd{and what it tells us about the fuelling of AGN in galaxies that lie along the main sequence} (Section~\ref{sec:discuss_sfgals}). 
\mupd{In Section~\ref{sec:discuss_massvssfr} we further explore whether stellar mass or SFR is the key parameter that determines the incidence of AGN in star-forming galaxies by comparing measurements of \Psar\ and the distribution of black hole accretion rates relative to the SFR.}
In Section~\ref{sec:discuss_belowMS} we discuss the processes that trigger and fuel AGN in galaxies with lower SFRs, below the main sequence of star formation at a given redshift. 
Section~\ref{sec:discuss_starbursts} briefly explores the possible role of mergers in the fuelling of AGN in starburst galaxies, with elevated SFRs compared to the main sequence. 

\subsection{The relationship between SFR and the incidence of AGN in main-sequence star-forming galaxies}
\label{sec:discuss_sfgals}

\mupd{In Section~\ref{sec:fagn_sfandqu} we found a linear correlation between the SFR of star-forming galaxies and both the fraction of star-forming galaxies that contain an AGN and the average specific accretion rates (\rone{see Figure~\ref{fig:fagn_vs_sfr}}), measured in a robust manner that accounts for the sensitivity limits of our X-ray imaging and stellar-mass-dependent selection biases.}
Thus, star-forming galaxies (at a fixed redshift) that are further along the main sequence---with higher stellar masses and thus higher SFRs---have correspondingly higher AGN fractions \mupd{and are typically growing their black holes at a higher rate.}
Furthermore, higher redshift star-forming galaxies (at fixed stellar mass) also have higher SFRs and a high incidence of AGN. 
Both the stellar-mass dependence and the redshift dependence of the incidence of AGN in star-forming galaxies appear to be explained, at least to first order, by the changes in the average SFRs with stellar mass and redshift.
Indeed, the \emph{differential} redshift evolution of AGN fractions in star-forming galaxies of different stellar masses, whereby the AGN fraction in higher mass galaxies appears to undergo a more rapid evolution than in lower mass galaxies (see figure~6 of \PaperII), can be partly explained by the differential evolution of SFRs at different stellar masses, whereby the average SFRs of high-mass galaxies evolve more rapidly with redshfit than in lower mass galaxies \citep[an effect related to the flattening of the main sequence at high masses and low redshifts e.g.][]{Schreiber15,Lee15}.

The observed correlation between the average SFR (at a given \Mstel\ and $z$) and the incidence of AGN appears broadly consistent with the well-established \emph{global} correlations between the total AGN accretion rate density and total SFR density over cosmic time \citep[e.g.][]{Boyle98b,Madau14,Aird15}.
However, our measurements demonstrate that a correlation between the SFR and the level of AGN activity exists within the star-forming galaxy population at a wide range of stellar masses and redshifts.
We note that our measurements are also consistent with the \emph{lack} of a direct correlation between AGN luminosity and the SFR in samples that are \emph{selected} based on the presence of an AGN \citep[e.g.][]{Azadi15,Stanley15}. 
The broad distribution of AGN accretion rates within a given sample of galaxies---tracked by our measurements of \Psar\ presented here as well as in \PaperII---blurs out any direct correlation between AGN luminosity and SFR in such samples.
\mupd{
However, by calculating the \emph{average} rate of accretion in a given sample of galaxies we summarize the broad distribution, average out any variability on galactic timescales, and a direct correlation is revealed \citep[see our measurements of \Avsar\ in the bottom panel of Figure~\ref{fig:fagn_vs_sfr}; see also][]{Chen13,Yang17}.}

\mupd{
Our measurements also quantify the \emph{fraction} of galaxies with significant levels of black hole growth at a given time, finding this quantity is also correlated with the SFR.
We have thus shown that an increase in the SFR of star-forming galaxies is associated with \emph{both} an increase in the average rate such galaxies are growing their black holes \emph{and} the fraction of the time they spend in active phases, hosting an AGN. 
These active phases dominate the average specific accretion rate, indicating that most black hole growth occurs in relatively short-lived periods of AGN activity.}

These correlations suggest that both AGN activity and star formation have a common physical origin in normal, main-sequence star-forming galaxies.
The SFR on galactic scales is ultimately determined by the availability of cold gas to form stars \citep[e.g.][]{Combes13,Saintonge13,Santini14}; if such gas is brought into the centres of galaxies it will also be accreted by the black hole and thus trigger periods of AGN activity. 
While the star formation will mainly proceed in a smooth and coherent manner across the galaxy, periods of AGN activity are likely to be much more stochastic due to the clumpiness or fragmentation of the gas distribution on scales relevant for the black hole, the irregular nature of the physical mechanisms that will drive such gas into the central regions (e.g. bars, major or minor mergers), and local feedback associated with the accretion onto the black hole itself. 
Such stochasticity is reflected in the broad distribution of accretion rates that we observe and the inferred variability of AGN activity \citep[see also][]{Mullaney12b,Aird13,Hickox14}.

\subsection{\mupd{Is AGN activity determined by stellar mass or SFR in star-forming galaxies?}}
\label{sec:discuss_massvssfr}

\begin{figure*}
\includegraphics[width=0.9\columnwidth,trim=0 0 10 10]{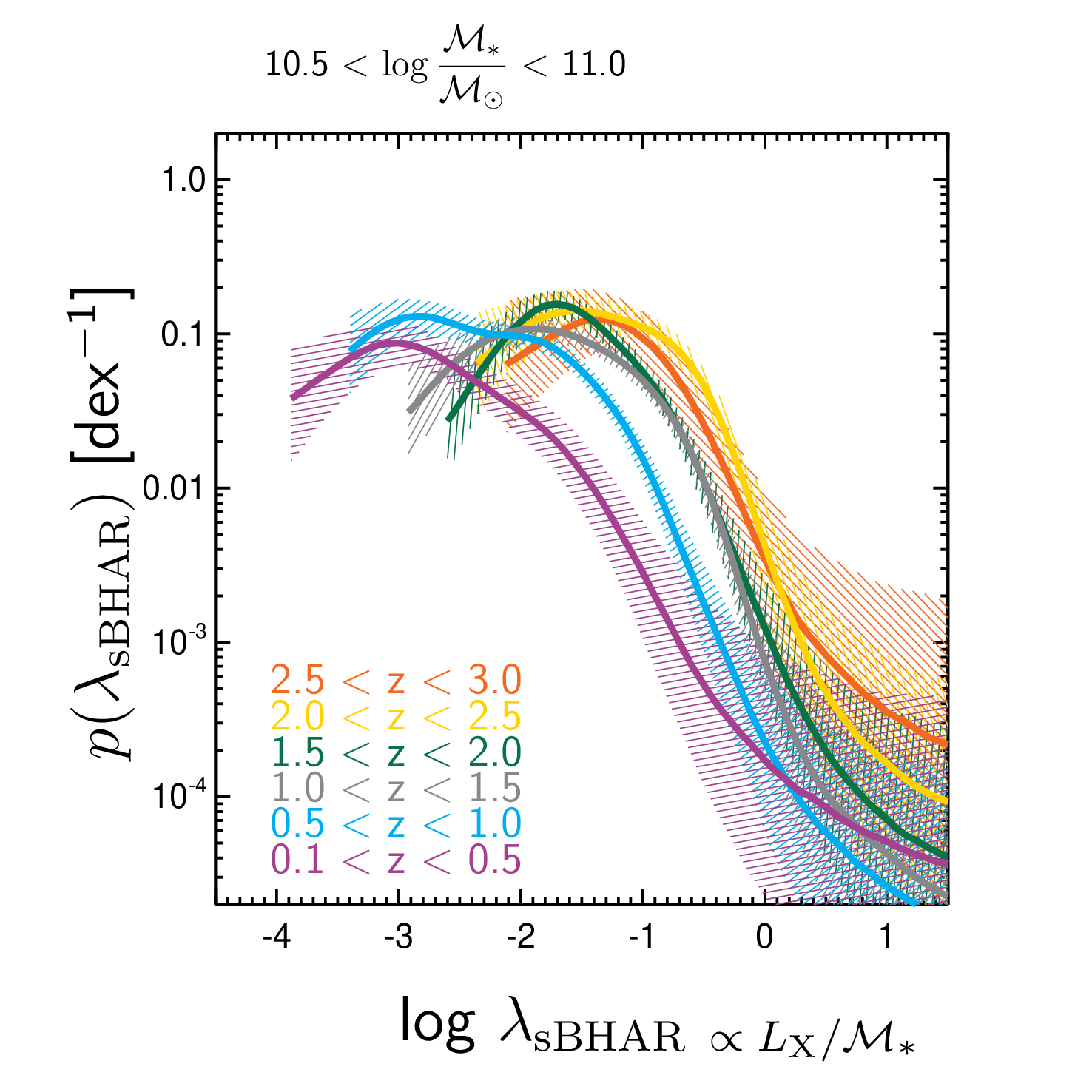}
\includegraphics[width=0.9\columnwidth,trim=0 0 10 10]{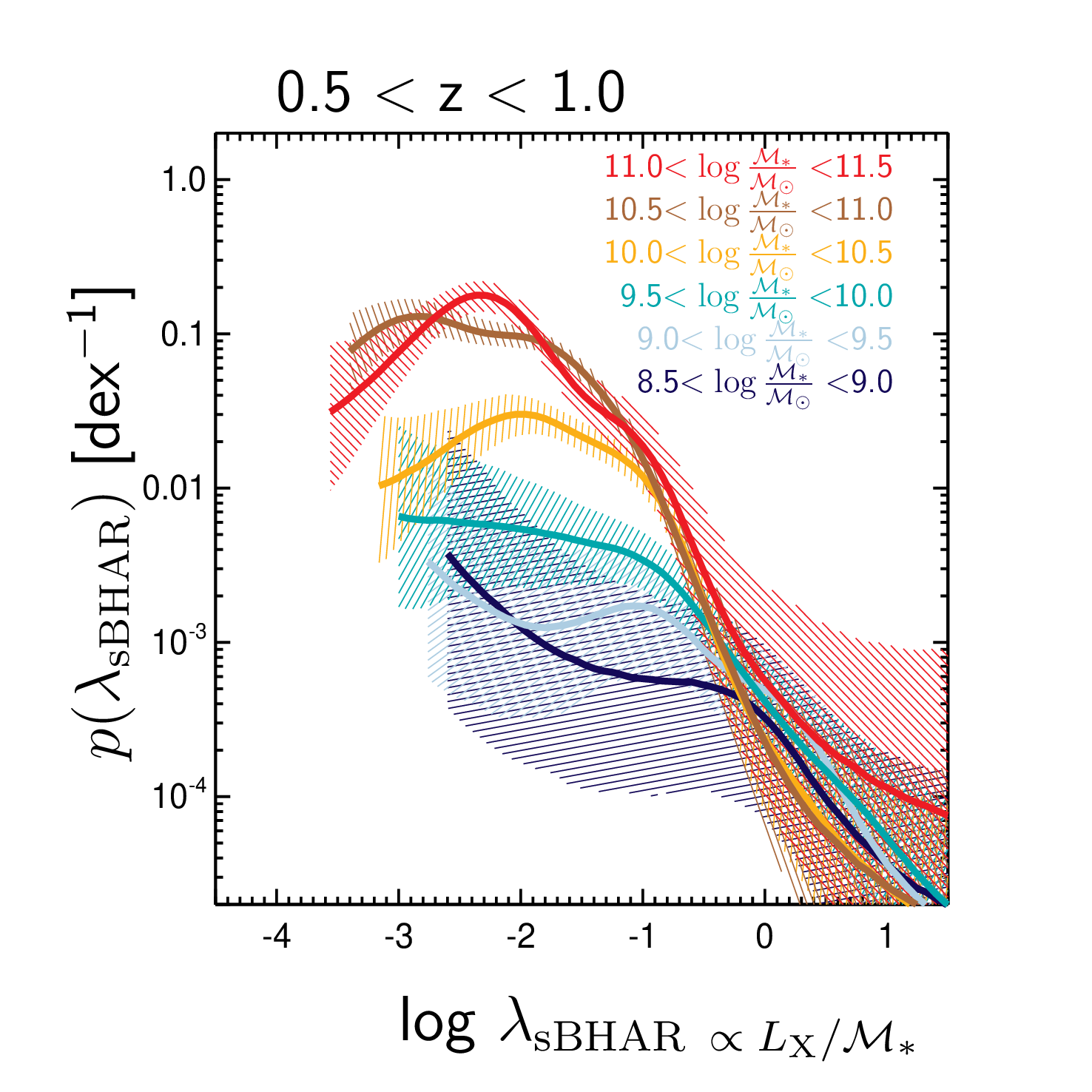}
\includegraphics[width=0.9\columnwidth,trim=0 10 10 10]{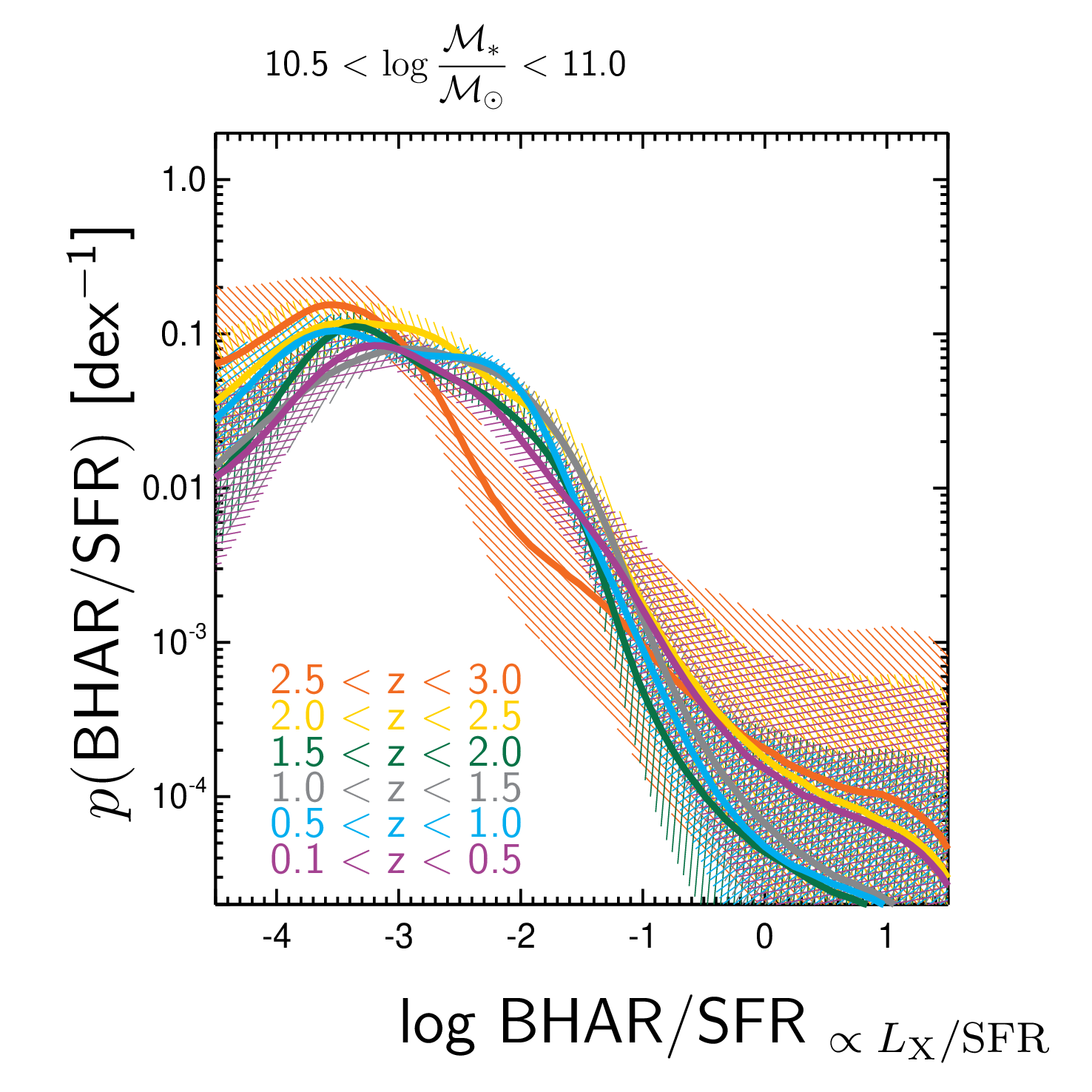}
\includegraphics[width=0.9\columnwidth,trim=0 10 10 10]{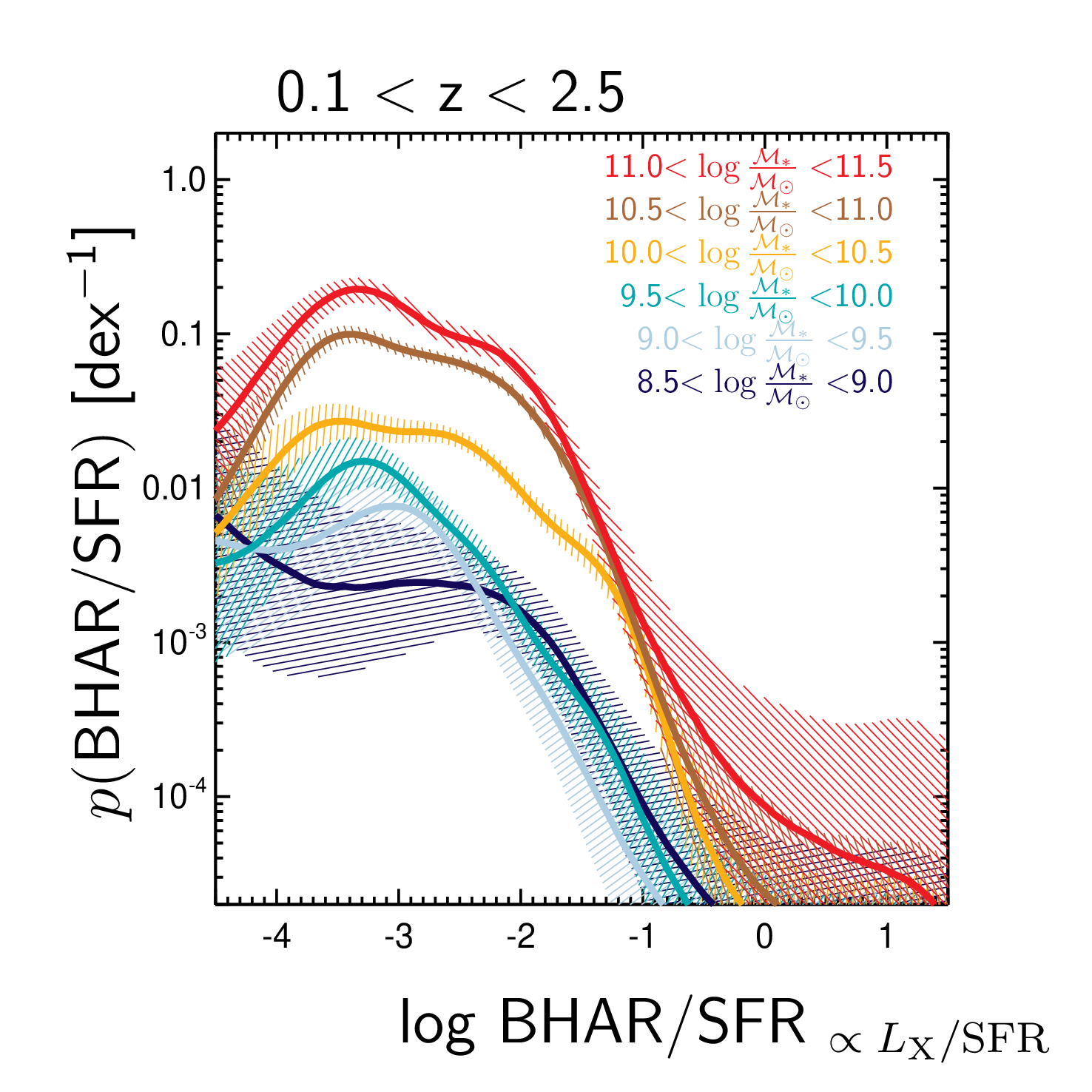}
\caption{
\upd{
\emph{Top row:} 
Examples of measurements of the probability distributions of specific black hole accretion rates (i.e. $L_\mathrm{X}$ normalized by the total stellar mass) in star-forming galaxies.
The top-left panel compares measurements at a fixed stellar mass but across a range of redshifts (as indicated by the colours), illustrating the strong evolution whereby \Psar\ shifts toward higher \Sar\ at higher redshift. 
The top-right panel shows the example of the stellar mass dependence of \Psar\ for a narrow redshift bin.
(See figures 4 and 5 of \PaperII\ for original measurements at all \Mstel\ and $z$).
\emph{Bottom row:}
New measurements of the probability distribution of BHAR/SFR, normalizing the X-ray luminosity by the SFR instead of stellar mass.
The bottom-left panel shows measurements at fixed stellar mass ($10.5<\log \mstel/\msun<11$) and at different redshifts (as indicated by the colours). 
We find that $p($BHAR/SFR$)$ has a consistent shape and normalization at fixed stellar mass at all redshifts (to $z\sim2.5$), indicating that the evolution in the distribution of AGN accretion rates in star-forming galaxies can be fully accounted for by the overall evolution of SFRs (cf. top-left panel).
The bottom-right panel presents measurements of $p($BHAR/SFR$)$ for samples of star-forming galaxies combined over a broad range in redshift ($0.1<z<2.5$) at a range of stellar masses (as indicated by the colours). 
While galaxies of all stellar masses exhibit a broad distribution of BHAR/SFR, reflecting variability in the levels of AGN activity, a significant stellar-mass dependence is still found, whereby the incidence of AGN is suppressed in lower mass star-forming galaxies. 
(See Appendix~\ref{app:bharsfr} for measurements of $p($BHAR/SFR$)$ over our full range of stellar mass and redshift).
}}
\label{fig:pbharsfr_examples}
\end{figure*}

Given the overall correlation between average SFR and stellar mass for star-forming galaxies on the main sequence, it is difficult to ascertain \emph{at a fixed redshift} whether the stellar mass or the SFR of galaxies is the key parameter that determines the incidence of AGN \citep[see also][]{Yang17,Yang18}.
However, the fact that a correlation between AGN fractions or average specific accretion rates and SFR persists over a wide range in redshift does suggest that the SFR---and the associated absolute amount of cold gas in a typical star-forming galaxy---is important. 
To further explore the importance of the absolute level of SFR in determining the level of AGN activity, we have repeated our measurements of the probability distributions of accretion rates in star-forming galaxies but normalizing the observed X-ray luminosity by the SFR instead of the total stellar mass. 
For each galaxy, we determine the absolute black hole accretion rate (BHAR), tracked by the X-ray luminosity and given by
\begin{equation}
\mathrm{BHAR} = \frac{\eta k_\mathrm{bol}}{c^2} L_\mathrm{X}
\end{equation}
where $k_\mathrm{bol}=25$ is our adopted bolometric correction, $\eta$ is the radiative efficiency (we adopt $\eta=0.1$), $c$ is the speed of light and the units are chosen such the BHAR is given in $\msun$~yr$^{-1}$.
Thus, BHAR/SFR is a unitless quantity, giving the ratio of the mass accreted onto the black hole relative to the mass formed into stars, and is directly proportional to the observed $L_\mathrm{X}/$SFR. 
As SFR and \Mstel\ are correlated in star-forming galaxies, normalizing \LX\ by the SFR
will still account for the broad stellar-mass-dependent selection bias whereby lower accretion rate sources are harder to detect in lower mass galaxies. We use our Bayesian methodology (see Section~\ref{sec:method} and the appendices of \PaperI\ and \PaperII\ for details) to combine the available X-ray data for all galaxies in a sample (both detections and non-detections at X-ray wavelengths) and recover estimates of the underlying probability distribution function, \Pbs.

Figure~\ref{fig:pbharsfr_examples} illustrates the key findings from our measurements of \Pbs\ (our full set of results in all stellar mass and redshift bins are provided in Appendix~\ref{app:bharsfr}) and compares them with our original measurements of \Psar\ (i.e. normalizing \LX\ by the stellar mass) from \PaperII.
\upd{
The left panels show measurements of \Psar\ and \Pbs\ in a single stellar mass bin at different redshifts. 
For \Psar\ (top-left panel), we see a strong redshift evolution whereby the overall distribution shifts toward higher \Sar\ at higher redshift.
However, we find that \Pbs\ (bottom-left panel) has a consistent shape and normalization across a wide range of redshifts.
}
The evolution of \Psar\ toward higher \Sar\ at higher redshifts and the associated evolution of the AGN fraction is \emph{fully} accounted for by the evolution in the average SFRs of star-forming galaxies, at least to $z\sim2.5$. 
Star-forming galaxies of a given stellar mass have a broad distribution of BHARs (spanning $\sim3-4$ orders of magnitude), but the level of accretion is determined purely by the SFR over a very wide range of redshifts. 
This redshift independence in \Pbs\ is found at all stellar masses (see Figure~\ref{fig:pbharsfr_massbins}). 

As \Pbs\ appears to be independent of redshift (at fixed \Mstel), we can combine our galaxy samples over the entire redshift range and recalculate \Pbs\ at each stellar mass. 
These measurements are shown in the bottom-right panel of Figure~\ref{fig:pbharsfr_examples}.
We note that, due to the flux limits of the NIR imaging, the lower mass galaxy samples do not span the full redshift range indicated in Figure~\ref{fig:pbharsfr_examples}, although this should not affect our measurements if we assume the redshift independence of \Pbs\ holds for all stellar masses.
While \Pbs\ has a broadly similar shape at all stellar masses, we find that there are significant differences in the overall normalization depending on stellar mass (cf. the stellar mass dependence of \Psar\ illustrated in the top-right panel of Figure~\ref{fig:pbharsfr_examples}).

\begin{figure}
\includegraphics[width=\columnwidth,trim=20 30 45 20]{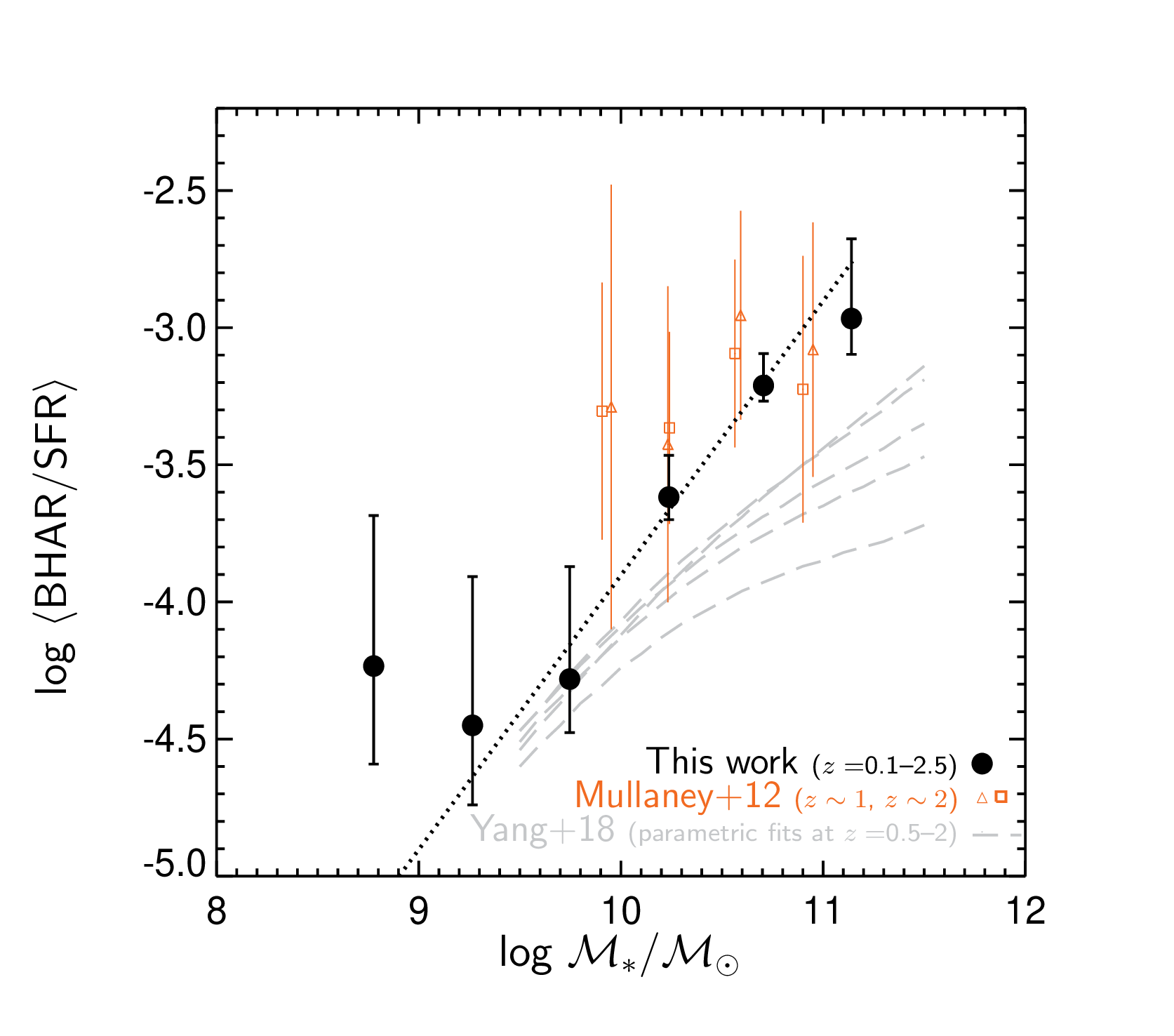}
\caption{
\mupd{
Measurements of $\langle$BHAR/SFR$\rangle$ as a function of stellar mass, derived from our measurements \Pbs\ for star-forming galaxies over a broad range in redshift, $0.1<z<2.5$ (black circles). 
We note that there is no significant evolution over this range of redshift based on our measurements of \Pbs\ in finer redshift bins (see Appendix~\ref{app:bharsfr}), indicating that $\langle$BHAR/SFR$\rangle$ remains constant at a fixed \Mstel\ out to at least $z\sim2.5$. 
\rone{The black dashed line is a linear 1:1 fit to our measurements (see Equation~\ref{eq:bharsfr_fit}).}
We compare our measurements (black circles) to previous estimates of the ratio of average black hole growth to average galaxy growth i.e.~$\langle$BHAR$\rangle$/$\langle$SFR$\rangle$ from \citet[red triangle and squares corresponding to measurements at $z\sim1$ and $z\sim2$, respectively]{Mullaney12b} and the parametric description for star-forming galaxies from \citet[grey dashed lines indicating different redshifts from $z=0.5$ to $z=2$]{Yang18}.
}
}
\label{fig:avbhartosfr_vs_mstel}
\end{figure}

\mupd{To further illustrate the stellar mass dependence of \Pbs, in Figure~\ref{fig:avbhartosfr_vs_mstel} we average over \Pbs\ to calculate the average ratio of black hole to stellar growth, $\langle$BHAR/SFR$\rangle$, over the $0.1<z<2.5$ redshift range.
Our measurements of $\langle$BHAR/SFR$\rangle$ (black circles in Figure~\ref{fig:avbhartosfr_vs_mstel}) 
range from $\sim3\times10^{-5}$ at $\mstel\sim10^{9}\msun$ to $10^{-3}$
at $\mstel\gtrsim10^{11}\msun$, indicating that more massive galaxies---on average---tend to have higher levels of black hole growth relative to the level of stellar growth than lower mass galaxies. 
\rone{We find that the dependence of $\langle$BHAR/SFR$\rangle$ on stellar mass can be described by a linear relation,}
\begin{equation}
\langle \mathrm{BHAR/SFR} \rangle =  \log \frac{\mstel}{10^{10}\msun} - (3.90\pm0.06)
\label{eq:bharsfr_fit}
\end{equation}
\rone{
which is shown by the black dotted line in Figure~\ref{fig:avbhartosfr_vs_mstel}. Allowing for a non-linear mass dependence in Equation~\ref{eq:bharsfr_fit} does not significantly improve the $\chi^2$ of the fit.
The lowest mass measurement appears to be a $\sim$2$\sigma$ outlier to the relation, although we cannot rule out a flattening of the relation at low stellar masses.
}
The redshift-independence of our measurements of \Pbs\ indicate that this relation is the same across a wide range of redshifts (since at least $z\sim2.5$), despite the significant evolution of both the typical SFRs of star-forming galaxies and the typical black hole growth rates over this period of cosmic time.
For high stellar mass galaxies ($\mstel\sim10^{11}\msun$), the ratio of black hole to galaxy stellar growth ($\sim10^{-3}$) is roughly consistent with the observed ratio of black hole to bulge mass in the local Universe \citep[e.g.][]{Marconi03,Kormendy13}.}
\rone{The linear slope of the relation seen in Figure~\ref{fig:avbhartosfr_vs_mstel} and described by Equation~\ref{eq:bharsfr_fit} indicates that the average level of black hole growth is strongly dependent on stellar mass.
Star-forming galaxies with $\mstel\sim10^{9}\msun$ will have a factor $\sim$25 lower absolute SFRs than galaxies with $\mstel\sim10^{11}\msun$ (as given by the main sequence) but are growing their black holes, on average, at a factor $\sim$100 less rapidly \emph{relative} to the SFR (and thus $\sim$2500 times less rapidly in an absolute sense).}

\rone{
To interpret the strong stellar-mass dependence of the \emph{average} black hole growth to stellar growth of star-forming galaxies, $\langle$BHAR/SFR$\rangle$, shown in Figure~\ref{fig:avbhartosfr_vs_mstel}, it is important to re-visit Figure~\ref{fig:pbharsfr_examples}.
This figure shows that
AGN may be triggered in star-forming galaxies of all stellar mass but the \emph{rate} at which they are triggered depends on the stellar mass, which is reflected in the \emph{normalization} of the probability distribution shown in the the bottom-right panel of Figure~\ref{fig:pbharsfr_examples}.
Once triggered, their accretion rates follow a broad distribution relative to the SFR of the galaxy. 
The SFR appears to set the distribution of accretion rates across a broad range in redshifts, i.e. the \emph{shape} of $p\mathrm{(BHAR/SFR)}$, consistent with the amount of cold gas determining both the SFR and the stochastically varying level of black hole accretion.  
However, the total stellar mass of a star-forming galaxy appears to have a strong effect on the triggering rate of these stochastic periods of cold gas accretion, i.e. sets the \emph{normalization} of $p\mathrm{(BHAR/SFR)}$. 
As discussed in \PaperII, this could be due the increased effects of stellar (supernova) feedback in lower mass galaxies that 
reduces the frequency with which gas clouds reach the galactic centre and trigger periods of AGN activity \citep[e.g.][]{Dekel86,Cole00,Hopkins14}.
It is this stellar-mass dependence of the \emph{triggering} rate that results in the strong stellar-mass dependence of $\langle$BHAR/SFR$\rangle$ seen in Figure~\ref{fig:avbhartosfr_vs_mstel}. Once triggered, lower mass galaxies produce AGN with comparable accretion rates to those within higher mass galaxies.}

\mupd{Our measurements of $\langle$BHAR/SFR$\rangle$ agree well with earlier work by \citet{Mullaney12b} who found a consistent ratio between average black hole growth and average galaxy growth across a range of stellar masses and redshifts; our measurements have smaller uncertainties and extend to lower stellar masses, revealing the strong stellar mass dependence. 
\citet{Yang18} also found a stellar mass dependence in the ratio of average black hole growth to average SFR, based on a parametric description of the distribution of accretion rates, and found a redshift dependence in the ratio over $z=0.5-4$. 
The absolute values they measure are significantly below our estimates at $\mstel\gtrsim10^{10}\msun$ (see Figure~\ref{fig:avbhartosfr_vs_mstel}).
However, we note a key difference in our analysis: we measure BHAR/SFR for each individual galaxy, derive the overall distribution, \Pbs, and average over this distribution to calculate $\langle$BHAR/SFR$\rangle$.
In contrast, \citet{Yang18} first measure the distribution of BHAR, take the average $\langle$BHAR$\rangle$, and compare to independent measurements of $\langle$SFR$\rangle$ based on the main sequence of star formation.
Such methodological differences, ultimately deriving a different quantity, may explain the offset in our results.
We also do not find significant redshift evolution (out to at least $z\sim2.5$) in our direct measurements of \Pbs\ or the $\langle$BHAR/SFR$\rangle$ derived therefrom at a fixed stellar mass, in contrast to the parametric derivation of $\langle$BHAR$\rangle$/$\langle$SFR$\rangle$ of \citet{Yang18} that exhibits a redshift dependence for star-forming galaxies between $z=0.5$ and $z=2$ (as indicated by the grey dashed lines in Figure~\ref{fig:avbhartosfr_vs_mstel}).
Our measurements reveal the detailed \emph{distribution} of BHAR/SFR and can thus provide more refined insights into the fuelling and triggering of AGN than the global averages.}

\subsection{The enhanced triggering of AGN in galaxies that lie below the main sequence}
\label{sec:discuss_belowMS}

We now consider galaxies with SFRs that place them below the main sequence of star formation and thus have lower levels of SFR than the bulk of star-forming galaxies at a given stellar mass and redshift. 
In Section~\ref{sec:fagn_vs_mainseq}, we found a significant increase (by a factor $\sim2-5$) in the AGN fraction in star-forming galaxies that lie below the main sequence (Sub-MS galaxies) at $0.5\lesssim z <2.5$, compared to in galaxies that lie on the main sequence (\rone{see Figure~\ref{fig:fduty_vs_sfr_onepanel_msbins}}). 
Moving to lower SFRs (relative to the main sequence) into the quiescent galaxy population, we find that the AGN fraction decreases again by a factor $\sim2-3$, falling yet further for quiescent galaxies with the lowest SFRs. 
However, the AGN fraction remains significantly above the broad correlation with SFR that is seen for normal (i.e. main-sequence) star-forming galaxies (\rone{black dashed, dot-dashed, and long-dashed lines in Figure~\ref{fig:fduty_vs_sfr_onepanel_msbins}}), suggesting that a different physical process may be responsible for supplying gas into the centres of such galaxies to trigger and fuel AGN activity. 

The observed increase in the AGN fraction for Sub-MS galaxies is consistent with a number of prior studies that have shown that AGN host galaxies have a lower average SFR and a broader overall distribution of SFRs than star-forming galaxies of equivalent stellar mass and redshift \citep[e.g.][]{Aird12,Shimizu15,Mullaney15b,Scholtz18}. 
The broader distribution of SFRs for AGN hosts means that the \emph{fraction} of galaxies hosting an AGN increases for sub-MS galaxies, as  quantified by our measurements,
while the \emph{majority} of AGN are still found in main-sequence galaxies \citep[see also][]{Rosario13,Georgakakis14}

In \PaperII, we proposed that stellar mass loss could provide the supply of low angular momentum gas needed to sustain AGN activity (at relatively  low accretion rates) in the quiescent galaxy population, in the absence of significant cold gas that likely dominates AGN fuelling in main-sequence galaxies \citep[see also][]{Ciotti07,Kauffmann09,Wang17}. 
However, stellar mass loss begins promptly ($\sim2-5$~Myr) after the formation of a stellar population, declining exponentially thereafter \citep[e.g.][]{Bruzual03}. 
Thus, stellar mass loss will also provide a gas supply in normal, main-sequence galaxies, fuelling some level of AGN activity, but is likely sub-dominant compared to the stochastic accretion of cold gas. 
Stellar mass loss alone cannot explain the \emph{increase} in the AGN fraction that is seen in Sub-MS galaxies as there will not be an \emph{increase} in stellar mass loss on the relevant timescales in such galaxies. 

An alternative explanation for the existence of AGN in both the Sub-MS and quiescent galaxy populations is that they are a relic of AGN activity that was triggered earlier in the lifetime of a galaxy, before the quenching of star formation throughout the galaxy began.
In such a model, AGN could be triggered and fuelled by the same cold gas supply that determines the SFR, but the star formation has subsequently quenched, transforming the galaxy into a Sub-MS or quiescent galaxy.
Indeed, the AGN itself could play a role in the quenching of star formation via AGN feedback processes \citep[e.g.][]{Bower06,Somerville08}.
However, the periods of luminous, radiatively efficient AGN activity in such a model must be stable on timescales that are longer than  the quenching of star formation and the subsequent transformation of the stellar population of the galaxy, estimated to take $\sim$100-500~Myr under the most rapid scenarios \citep[e.g.][]{Wild09,Barro13}.
The broad distribution of AGN accretion rates in Sub-MS and quiescent galaxies, as found in our measurements, as well as independent observational evidence \citep[e.g.][]{Schawinski15} and theoretical arguments \citep[e.g.][]{King15} suggest that radiatively efficient accretion is unlikely to be stable on timescales of $\gtrsim0.1$~Myr. 
Thus, it is unlikely that the increase in the AGN fraction in Sub-MS galaxies can be directly associated with AGN feedback.

Given the short timescales of AGN activity, our results suggest that the rate at which AGN are \emph{triggered} increases for galaxies that lie just below the main sequence. 
It is unclear what physical mechanism or process results in this increased rate of triggering. 
One possibility is that the increase in AGN triggering is related to the build up of the central bulges of galaxies, observed as an increase (at fixed total stellar mass) in the central density for galaxies that lie below the main sequence \citep[e.g.][]{Wuyts11a,Cheung12,Mendez13}. 
Thus, the processes that drive gas into the centres of galaxies, forming stars and increasing the central mass density, may also increase the rate at which AGN are triggered. 
Indeed, recent studies have revealed an association between the compactness of star-forming galaxies and a high incidence of AGN activity \citep[e.g.][]{Barro13,Kocevski17} that appears consistent with this proposal.

\subsection{The role of mergers in the fuelling of AGN in starburst galaxies} 
\label{sec:discuss_starbursts}

Finally, we discuss the fuelling of AGN in starburst galaxies, with SFRs that are enhanced relative to the bulk of star-forming galaxies at a given redshift i.e. are above the main sequence. 
Our measurements show that \Fduty, \Fbright\ and \Avsar\ also increase in such galaxies compared to within main sequence galaxies (see Figure~\ref{fig:fduty_vs_sfr_onepanel_msbins}), indicating an increase in AGN activity that is associated with the increase in SFR. 
This enhancement can be examined in more detail using our measurements of \Psar\ (Figure~\ref{fig:pledd_fivebins}), which show an excess at the highest accretion rates ($\sar\sim0.1-1$) in the starburst galaxy samples, indicating that the most rapid accretion events are enhanced within starburst galaxies. 
These results are consistent with the findings of \citet{Bernhard16}, who found an association between the highest \Sar\ AGN and starburst-level SFRs, and \citet{Rodighiero15} who find that the average specific accretion rates (measured via stacking) are higher in starburst galaxy samples. 

The enhanced SFRs appear to follow the same overall correlations between incidence of AGN and the SFR identified throughout the star-forming galaxy population in Section~\ref{sec:fagn_sfandqu}.
Thus, the increase in the incidence of high-\Sar\ events may simply reflect the higher SFRs of the starburst galaxies.
AGN in starburst galaxies could be fuelled by the stochastic accretion of cold gas, as in normal star-forming galaxies, and the increased fraction simply reflects a higher abundance of gas that also produces the elevated SFRs in such galaxies. 
However, the redshift evolution of \Fduty, \Fbright\ and \Avsar\ in starburst galaxies is weaker than in the rest of the galaxy population, changing by a factor $\lesssim3$ between $z\sim0.3$ and $z\sim2$ (see Figure~\ref{fig:fduty_vs_z_fivebins}), suggesting a different physical mechanism may be responsible for enhancing the incidence of AGN in such galaxies, particularly at lower redshifts.

It is now well established that major mergers play a primary role in enhancing SFRs to produce the population of starburst galaxies \citep[e.g.][]{Sanders96,Lonsdale06}, particularly at lower redshifts where the typical SFRs of galaxies are much lower \citep[e.g.][]{Rodighiero11}. 
Such a violent process is also expected to drive significant quantities of gas into the centres of galaxies and trigger substantial AGN activity \citep[e.g.][]{Hopkins06b}. 
In addition, the merger rate (at a given stellar mass) evolves mildly with redshift, by a factor $<2$ between $z\sim0$ and $z\sim2$ \citep[e.g.][]{Lotz11}, similar to our AGN fraction in starburst galaxies.
We thus propose that the enhanced AGN fraction in starburst galaxies---along with the shift in their distribution of accretion rates toward higher values and the mild redshift evolution---may be predominantly due to major mergers.

We note that major mergers are not \emph{required} to trigger an AGN or even a high-luminosity AGN. 
Our results simply indicate a higher incidence of AGN in starburst galaxies that appears to be associated with an increased role of mergers.
The bulk of observed AGN are found in normal star-forming galaxies and have a broad distribution of accretion rates, reflecting stochasticity in the accretion process that can still produce high-\Sar\ (and thus high-luminosity) AGN that are not necessarily associated with mergers \citep[see also][]{Mechtley16,Villforth17,Goulding18}.

\section{Summary and Conclusions}
\label{sec:summary}

This paper presents robust measurements \mupd{of the incidence of AGN in galaxies as a function of their SFR.}
\upd{
We start from a sample of $\sim120,000$ near-infrared selected galaxies (from the CANDELS and UltraVISTA surveys) out to $z\sim4$ within which we identify populations of galaxies with different stellar masses, redshifts and SFRs.
We then use deep \textit{Chandra} imaging to extract X-ray data for \emph{all} galaxies in a given sample, which we combine using an advanced Bayesian methodology to determine the underlying 
probability distribution function of specific accretion rates, \Psar.
\mupd{From these probability distributions, we derive robust AGN fractions that account for both the varying sensitivity limits of the X-ray imaging and the stellar-mass-dependent selection bias, along with estimates of the average specific black hole accretion rates for a given galaxy sample.}
Initially, we consider samples of star-forming galaxies at different stellar masses---probing changes in the average SFRs \emph{along} the main sequence of star formation---and compare to quiescent galaxy populations (Section~\ref{sec:fagn_sfandqu}). 
We then divide the galaxy sample (at fixed \Mstel) according to their SFRs relative to the main sequence of star formation i.e.~probing \emph{across} and below the main sequence (Section~\ref{sec:fagn_vs_mainseq}).
}
Our main conclusions are as follows:
\begin{enumerate}
\item
Within star-forming galaxies, we find a linear correlation between the \mupd{incidence of AGN (characterised by both the AGN fraction to different limits and the average specific accretion rate)} and the average SFR---moving \emph{along} the main sequence of star formation---that persists across a wide range in both stellar mass ($8.5<\log\mstel/\msun<11.5$) and redshift (to at least $z\sim2.5$).
This correlation indicates that the SFR is the key observable property that determines AGN activity in the bulk of star-forming galaxies and suggests that both have a common physical origin e.g. both are determined by the amount of cold gas in a galaxy.

\item
We also show that star-forming galaxies of a fixed stellar mass have a consistent distribution of accretion rates relative to the total SFR in a galaxy (i.e. $L_\mathrm{X}/$SFR) over a wide range in redshift (to $z\sim2.5$), but the normalization of the distribution function depends on stellar mass. 
Thus, while the SFR determines the \emph{distribution} of AGN accretion rates in star-forming galaxies, the \emph{triggering} rate appears to be suppressed at lower stellar masses. 
\mmupd{On average, lower mass galaxies are growing their black holes at a lower rate relative to their total stellar mass, compared to higher mass galaxies.}

\item
We find that the AGN fraction in quiescent galaxies is significantly higher than in star-forming galaxies with equivalent SFRs (but with different stellar masses and redshifts), indicating that another physical mechanism---such as stellar mass loss---may fuel ongoing AGN activity within such quiescent galaxies.
\upd{We also note that the AGN fraction may be more closely correlated with the sSFR for quiescent galaxies, in contrast to the absolute value of SFR that appears to determine the AGN fraction in main sequence star-forming galaxies.} 

\item
We divide our galaxy sample in a more refined manner, according to their SFRs relative to the main sequence of star formation (at fixed stellar mass) and thus probing \emph{across} the main sequence.
Within the star-forming galaxy population, the AGN fraction is \emph{lowest} for galaxies that lie \emph{on} the main sequence.
These galaxies correspond to the majority ($\sim$ 70 percent) of star-forming galaxies and thus, despite the lower AGN \emph{fraction} still host the majority of AGN. 

\item
The AGN fraction is a factor $\sim2-5$ \emph{higher} in star-forming galaxies with \emph{lower} SFRs, placing them below the main sequence.
This increase indicates that the triggering of AGN is enhanced in these sub-main sequence galaxies.
The physical mechanism that drives this excess remains unclear \mmupd{but may be related to the build up of a central stellar bulge.}

\item
The AGN fraction and the average specific accretion rate is enhanced in starburst galaxies, which have SFRs that place them above the main sequence at a given redshift. 
In addition, the incidence of AGN in these starburst galaxies appears to have a milder evolution with redshift, changing by a factor $\lesssim 3$ between $z\sim0.3$ and $z\sim2$, suggesting that major galaxy mergers preferentially fuel the high accretion rate AGN found in starburst galaxies.  

\end{enumerate}

Our work shows that, while the bulk of AGN activity tracks star formation over cosmic time, a number of diverse physical processes are responsible for triggering and fuelling AGN across the galaxy population.
Our measurements of \Psar, both along and across the main sequence as presented here, provide crucial observational constraints that can be compared to large-scale cosmological simulations of galaxy and black hole evolution or can be used to accurately populate galaxies with AGN in observationally motivated models.

\section*{acknowledgements}
We thank the referee for helpful comments that improved the clarity of this paper.
We acknowledge helpful conversations with Jeremy Bradford and Meg Urry.
JA acknowledges support from an STFC Ernest Rutherford Fellowship, grant code: ST/P004172/1.  
AG acknowledges the {\sc thales} project 383549 that is jointly funded by the European Union  and the  Greek Government  in  the framework  of the  programme ``Education and lifelong learning''. 
This work is based in part on observations taken by the CANDELS Multi-Cycle Treasury Program and the 3D-HST Treasury Program (GO 12177 and 12328) with the NASA/ESA HST, which is operated by the Association of Universities for Research in Astronomy, Inc., under NASA contract NAS5-26555.
Also based in part on the K$_{s}$-selected catalogue of the COSMOS/UltraVISTA field from \citet{Muzzin13}, which contains PSF-matched photometry in 30 photometric bands and includes the available \textit{GALEX} \citep{Martin05}, CFHT/Subaru \citep{Capak07}, UltraVISTA \citep{McCracken12}, S-COSMOS \citep{Sanders07}, and zCOSMOS \citep{Lilly09} datasets.
The scientific results reported in this article are based to a significant degree on observations made by the \textit{Chandra} X-ray Observatory.

\appendix
\section{Measurements of the probability distribution of BHAR/SFR as a function of stellar mass and redshift}
\label{app:bharsfr}

\begin{figure*}
\includegraphics[width=\textwidth,trim=0 0 0 0]{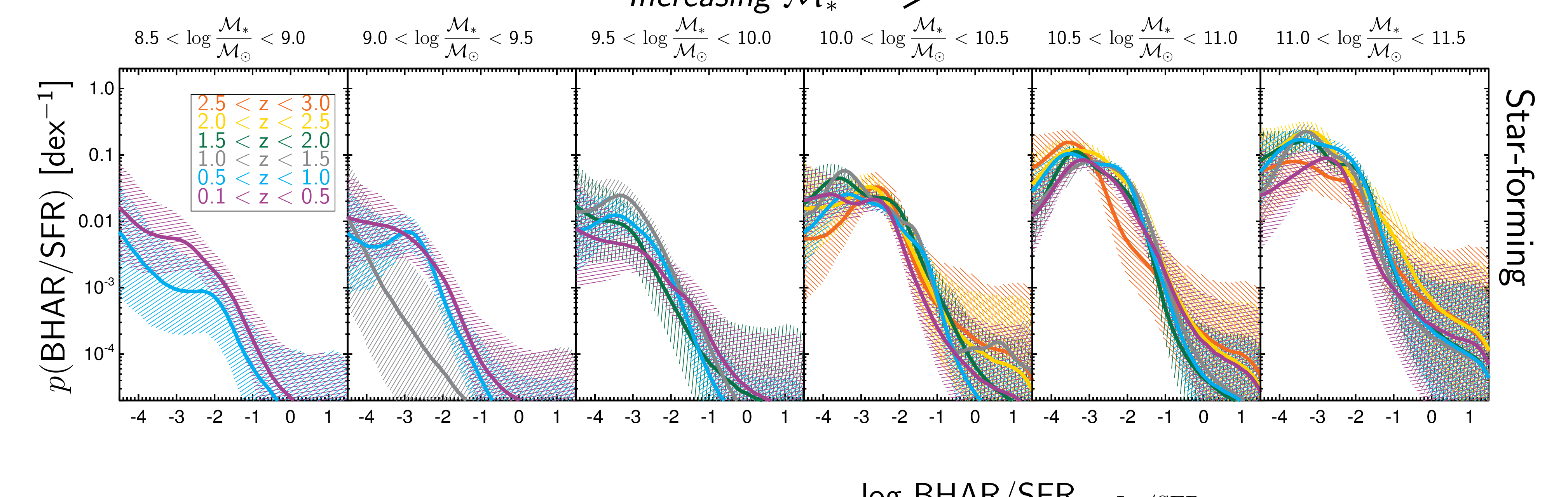}
\caption{
Measurements of $p($BHAR/SFR$)$ for star-forming galaxies, comparing different redshifts (as indicated by the colours) at fixed stellar mass. We find that $p($BHAR/SFR$)$ is consistent at fixed stellar mass over a wide range in redshift.
}  
\label{fig:pbharsfr_massbins}
\includegraphics[width=\textwidth,trim= 0 0 0 -20]{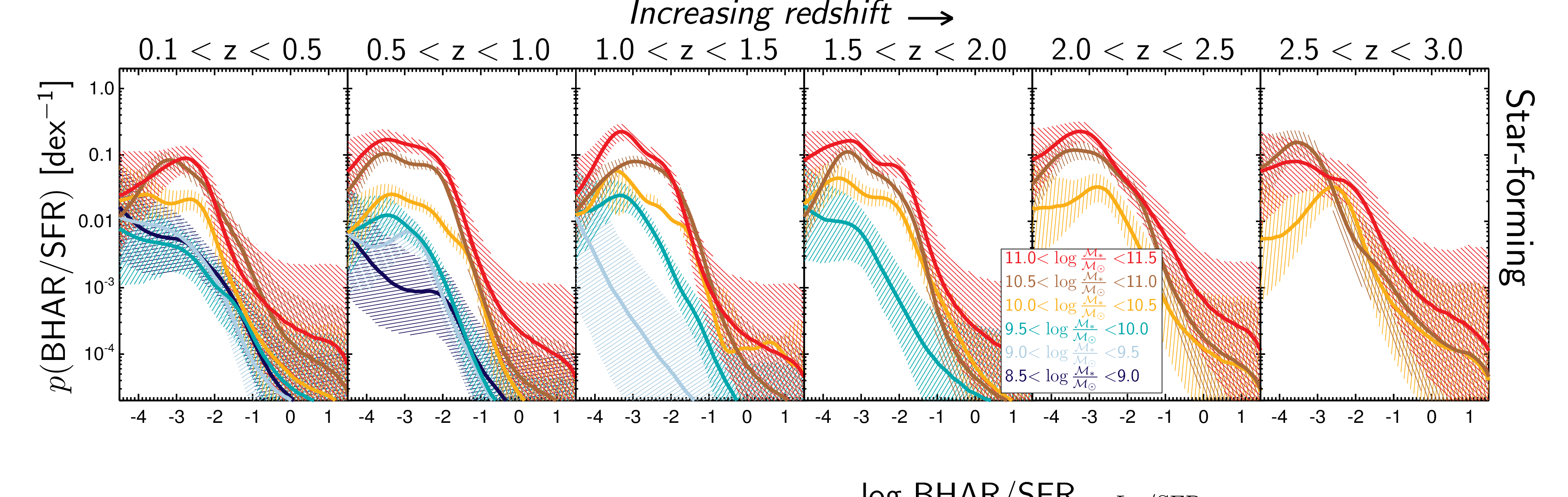}
\caption{
Measurements of $p($BHAR/SFR$)$ for star-forming galaxies, now comparing different stellar mass ranges (as indicated by the colours) at fixed redshift. 
A stellar mass dependence, whereby $p($BHAR/SFR$)$ is suppressed in lower mass galaxies, is found in the star-forming galaxy samples at all redshifts. 
}
\label{fig:pbharsfr_zbins}
\end{figure*}

This appendix presents measurements of the probability distribution function of BHAR/SFR (essentially, the observed X-ray luminosity normalized relative to the SFR of the galaxy) as a function of stellar mass and redshift for our star-forming galaxy samples across our full range of redshift and stellar mass.
Figure~\ref{fig:pbharsfr_massbins} shows \Pbs\ at fixed stellar masses for different redshifts (indicated by the colours); we find that \Pbs\ at a given stellar mass is consistent over a broad range of redshifts, indicating that the SFR determines the level of accretion. 
Figure~\ref{fig:pbharsfr_zbins} compares measurements at a given redshift for different stellar masses; a stellar mass dependence is found in star-forming galaxies at all redshifts, whereby the normalization of \Pbs\ is lower at lower \Mstel. 
We note that the same results are obtained if we restrict our analysis to just those star-forming galaxies that are \emph{on} the main sequence of star formation (i.e.~with SFRs within $\pm$0.4~dex of the main sequence), which dominate the star-forming population.

\label{lastpage}
\end{document}